%% file: main.tex
\documentclass[11pt]{article}

\input{header.tex}

\makeatletter
\def\blfootnote{\xdef\@thefnmark{}\@footnotetext}
\makeatother

\begin{document}

\title{Fully Dynamic Euclidean Bi-Chromatic Matching in Sublinear Update Time}
\author{Gramoz Goranci \thanks{Faculty of Computer Science, University of Vienna, Austria.} \and Peter Kiss \thanks{Faculty of Computer Science, University of Vienna, Austria. This research was funded in whole or in part by the Austrian Science Fund (FWF) 10.55776/ESP6088024} \and Neel Patel \thanks{University of Southern California, CA, US.} \and Martin P. Seybold \footnotemark[1] \and Eva Szilagyi \thanks{Faculty of Computer Science, UniVie Doctoral School Computer Science DoCS, University
of Vienna, Austria.} \and Da Wei Zheng \thanks{University of Illinois at Urbana-Champaign, IL, US.}}

\maketitle

\pagenumbering{gobble}

\input{abstract.tex}

\pagenumbering{arabic}
\newpage
\input{intro.tex}



\bibliographystyle{plain}
\bibliography{arxiv version/references}


\appendix
\end{document}

%% file: header.tex
\usepackage{fullpage}

\usepackage{xcolor}
\usepackage{pdfsync}
\usepackage{commath}

\usepackage{comment}

\usepackage{algpseudocode}
\usepackage{algorithm}

\newcommand{\match}{\textsc{Match}}
\newcommand{\staticmatch}{\textsc{Static-Matching}}
\newcommand{\cross}{\text{\textbf{Cross}}}

\renewcommand{\epsilon}{\varepsilon}

\def\showauthornotes{1}

\def\showdraftbox{0}

\input{macros}

\allowdisplaybreaks

\usepackage[backend=biber]{biblatex}
\renewcommand\leq{\leqslant}
\renewcommand\geq{\geqslant}

\newcommand{\eps}{\varepsilon}

\renewcommand{\setminus}{\backslash}



%% file: macros.tex
\usepackage{fancybox}

\usepackage{amsmath,amssymb,amsthm,amstext,amsfonts,bbm,algorithm,algorithmicx,graphicx,xspace,nicefrac}
\usepackage{color,stmaryrd,enumerate,latexsym,bm,amsfonts,wrapfig,verbatim,tabularx,textcomp,
subfig}

\usepackage{amsfonts}
\usepackage{comment} 
\usepackage{epsfig} 
\usepackage{latexsym,nicefrac,bbm}
\usepackage{xspace}
\usepackage{color,fancybox,graphicx,url}
\usepackage{enumitem}
\usepackage{booktabs}
\usepackage{commath}
\usepackage{mdframed}
\usepackage{pdfsync}

\usepackage{thm-restate}

\usepackage{fancybox}
\newenvironment{fminipage}%
  {\begin{Sbox}\begin{minipage}}%
  {\end{minipage}\end{Sbox}\fbox{\TheSbox}}

\definecolor{ForestGreen}{rgb}{0.1333,0.5451,0.1333}
\definecolor{DarkRed}{rgb}{0.8,0,0}
\definecolor{Red}{rgb}{1,0,0}
\usepackage[linktocpage=true,colorlinks,
linkcolor=DarkRed,citecolor=ForestGreen,
bookmarks,bookmarksopen,bookmarksnumbered]
 {hyperref}

\usepackage{cleveref}
\usepackage{mathtools}
\usepackage{thmtools}

\declaretheorem[numberwithin=section]{theorem}
\declaretheorem[numberlike=theorem]{lemma}

\declaretheorem[numberlike=theorem]{claim}

\theoremstyle{remark}
\newtheorem*{rem*}{Remark}

\newcommand{\marginlabel}[1]%
{\mbox{}\marginpar{\it{\raggedleft\hspace{0pt}#1}}}

\definecolor{Mygray}{gray}{0.8}

\ifcsname ifcommentflag\endcsname\else
\expandafter\let\csname ifcommentflag\expandafter\endcsname
\csname iffalse\endcsname
\fi

\ifnum\showauthornotes=1
\newcommand{\todo}[1]{\colorbox{Mygray}{\color{red}\parbox{\textwidth}{#1}}}
\else
\newcommand{\todo}[1]{}
\fi

\ifnum\showauthornotes=1
\newcommand{\Authornote}[2]{{\sf\small\color{red}{[#1: #2]}}}
\newcommand{\Authoredit}[2]{{\sf\small\color{red}{[#1]}\color{blue}{#2}}}
\newcommand{\Authorcomment}[2]{{\sf \small\color{gray}{[#1: #2]}}}
\newcommand{\Authorfnote}[2]{\footnote{\color{red}{#1: #2}}}
\newcommand{\Authorfixme}[1]{\Authornote{#1}{\textbf{??}}}
\newcommand{\Authormarginmark}[1]{\marginpar{\textcolor{red}{\fbox{
#1:!}}}}
\else
\newcommand{\Authornote}[2]{}
\newcommand{\Authoredit}[2]{}
\newcommand{\Authorcomment}[2]{}
\newcommand{\Authorfnote}[2]{}
\newcommand{\Authorfixme}[1]{}
\newcommand{\Authormarginmark}[1]{}
\fi

\newlength{\pgmtab}  
 {
	\begin{enumerate}}{\end{enumerate}}

\def\qedsketch{\ifmmode\Box\else{\unskip\nobreak\hfil
\penalty50\hskip1em\null\nobreak\hfil$\Box$
\parfillskip=0pt\finalhyphendemerits=0\endgraf}\fi}

\newlength{\tpush}
\newcommand{\handout}[5]{
   \noindent
   \begin{center}
   \framebox{ \vbox{ \hbox to \textwidth { {\bf \coursenum\ :\  \coursename} \hfill #5 }
       \vspace{3mm}
       \hbox to \textwidth { {\Large \hfill #2  \hfill} }
       \vspace{1mm}
       \hbox to \textwidth { {\it #3 \hfill #4} }
     }
   }
   \end{center}
   \vspace*{4mm}
   \newcommand{\lecturenum}{#1}
   \addcontentsline{toc}{chapter}{Lecture #1 -- #2}
}

\ifnum\showdraftbox=1

\else

\fi



%% file: abstract.tex
\begin{abstract}
We consider the Euclidean bi-chromatic matching problem in the dynamic setting, where the goal is to efficiently process point insertions and deletions while maintaining a high-quality solution. Computing the minimum cost bi-chromatic matching is one of the core problems in geometric optimization that has found many applications, most notably in estimating Wasserstein distance between two distributions. In this work, we present the first fully dynamic algorithm for Euclidean bi-chromatic matching with sub-linear update time. For any fixed $\varepsilon > 0$, our algorithm achieves $O(1/\varepsilon)$-approximation and handles updates in $O(n^{\varepsilon})$ time.  Our experiments show that our algorithm enables effective monitoring of the distributional drift in the Wasserstein distance on real and synthetic data sets, while outperforming the runtime of baseline approximations by orders of magnitudes.
\end{abstract}


%% file: intro.tex

\section{Introduction}

We consider the Euclidean bi-chromatic matching problem, one of the most fundamental optimization problems in low-dimensional Euclidean geometry.
Given a set $A$ of $n$ red points in $\mathbb{R}^d$ and a set $B$ of $n$ blue points in $\mathbb{R}^d$, the goal is to compute a bijection $\mu : A \to B$ such that $\sum_{a \in A} \lVert a-\mu(a)\rVert_2$ is minimized. A key application of minimum-cost bi-chromatic matching in machine learning is estimating the $1$-Wasserstein distance of two spatial distributions $\nu$ and $\nu'$, whose supports satisfy $S_\nu,~S_{\nu'} \subseteq \mathbb{R}^d$. This distance, also known as Euclidean Earth Mover's Distance (EMD), Kantorovich-Rubinstein metric, or Mallow's distance, is defined as
\[
W_1(\nu, \nu')= \underset{P\in\Gamma(\nu,\nu')}{\inf} \int_{S_\nu \times S_{\nu'}} \lVert a - b \rVert_2 ~dP(a,b) \quad,
\]
were $\Gamma(\nu,\nu')$ denotes the set of all probability measures on $S_\nu\times S_{\nu'}$ with marginal distributions $\nu$ and $\nu'$.

 Estimating the $1$-Wasserstein distance measure, or the Euclidean bi-chromatic matching problem, has been extensively studied across a wide range of disciplines, including machine learning \cite{LiuGS18, LiuYZZ17,LiuGS18, CaoM0JST19, BalajiCF20,TolstikhinBGS18}, computer vision \cite{RubnerTG00}, statistics \cite{PanaretosZ19} and economics \cite{galichon2018optimal}. This broad interest has led to different problem varaiants, e.g., the point reweighted setting, and the development of algorithms with high-accuracy approximation guarantees or those that only estimate the solution size without explicitly outputting the underlying matching~(for a more detailed discussion, please see Appendix~\ref{app:sec:related-works}). 

 In this work, we consider the Euclidean bi-chromatic matching in the \emph{dynamic setting}, where the input undergoes \emph{updates} such as insertions or deletions of pairs of points. The goal is to process these updates as \emph{fast} as possible while maintaining a matching that achieves a small, provable approximation ratio compared to the optimal matching at any given time. 
 
 The dynamic setting is particularly relevant in the statistical estimation of the $1$-Wasserstein distance~\cite{GattaniRS23, BeugnotGG021, LiuGS18, PanaretosZ19}, when the closed-form expressions for $\nu$ and $\nu'$ are unknown. In such cases, one can efficiently draw independent and identically distributed (i.i.d.) samples from both distributions. 
The resulting empirical distribution $\nu_n$ and $\nu'_n$, assign a uniform probability mass of $1/n$ to each sample in $A$ and $B$, respectively.
It is well known that $W_1(\nu_n,\nu'_n)$ converges to $W_1(\nu,\nu')$, as $n \to \infty$. 
Rather than computing a matching from scratch after new samples are drawn, a dynamic algorithm guarantees efficient maintenance of accurate estimates for $\min_{\mu} \frac{1}{n} \sum_{a \in A} \lVert a-\mu(a)\rVert_2$ under such changes. Further real-world applications of dynamic geometric matching include measuring the similarity between evolving data sets~\cite{Alvarez-MelisF20}, measuring the change in patient data (such as MRI images) over a long period of time~\cite{GramfortPC15, JanatiBTCG19}, or applications of matching (or EMD) as a metric in time series analysis~\cite{ChengAHM21}. Despite the fundamental importance of this problem, dynamic geometric matching has only recently been considered in \cite{xu2024novel} and the existing algorithms cannot break the linear barrier on the update time.

\subsection{Main Contributions}
 We obtain the first sub-linear dynamic algorithm for Euclidean bi-chromatic bipartite matching. Specifically, our algorithm achieves $O(1)$ approximation in sub-linear in $n$ update time. The exact guarantees of our result are summarized in the theorem below.

\begin{restatable}{theorem}{main}
\label{main}

    For any $0 < \eps \leq 1$, there exists a fully dynamic algorithm that maintains an expected $O(1/\epsilon)$-approximate solution to the Euclidean bi-chromatic matching problem defined on the point-sets $A,B \subset \mathbb{R}^2$, $|A| = |B|$, while pairs of points are inserted into and deleted from $A,B$ in $O(n^{\epsilon} \cdot \epsilon^{-1})$ worst-case update time. Here, $n$ denotes the maximum size of $A$ throughout the update sequence. 
    
\end{restatable}



In the static setting, the Euclidean bi-chromatic matching problems admits known $(1+\epsilon)$-approximation algorithms that run in near-optimal running time 
\cite{swat/AgarwalRSS22,jacm/RaghvendraA20}. In comparison, the $O(1/\epsilon)$ approximation guarantee in Theorem~\ref{main} may initially seem less competitive. However, as we show in Appendix~\ref{app:hardness}~(Theorem~\ref{thm:hardness}), a simple lower bound construction reveals that, in the dynamic setting (even if restricted to only point insertions or deletions), there is no algorithm that simultaneously achieves $(2-\delta)$-approximation and runs in sub-linear update time, for any $\delta > 0$. This in turn implies that no non-trivial dynamic algorithm can maintain a solution to the problem with an arbitrarily tight approximation ratio.



While the matching problem on graphs has received significant attention in the dynamic model (see \cite{AzarmehrBR24} for a complete list of works), the only prior known algorithm for the geometric version of the problem was presented by \cite{xu2024novel} with update time $O(n)$. Notably, even disregarding edge weights, the maintenance of a perfect matching in the graph setting requires $\Omega(n)$ update time under well accepted hardness conjectures \cite{HenzingerKNS15,Dahlgaard16}. On the other hand, there is a long line of work focusing on maintaining an $O(1)$-approximate matching in near optimal update time in dynamic graphs.

Furthermore, dynamic graph matching algorithms focus on singular edge updates. As in the dynamic Euclidean bi-chromatic matching problem we have to work with updates affecting $\Omega(n)$ point-wise distances, in order to obtain our results, we need techniques which are new to dynamic literature and heavily exploit the guarantees of the geometric setting.

To complement our primary contribution~(Theorem~\ref{main}), we present experimental evaluations of the performance of our dynamic algorithm for estimating the $1$-Wasserstein distance in contrast to static state of the art approximate algorithms. Our experimental results reveal that in the dynamic setting the total running time of our algorithm over a series of updates outperforms periodic static recomputation by orders of magnitudes. Furthermore, we find that the approximation ratio obtained by our algorithm on real-world datasets is smaller than that of our theoretical guarantees and almost matches that of static algorithms. Our results confirm the expected approximation ratio drop-off and running time decrease we would expect with the reduction of the $\epsilon$ parameter.

\subsection{Technical Overview}

In vein of related works, we discuss algorithms and analysis for $d=2$ to simplify the presentation, though our theorems and implementation extend to constant $d>2$.

Our dynamic algorithm builds on the static algorithm of \cite{AV04}. On a very high level, the algorithm partitions the input points into a series of nested grid cells. The nested grid is constructed in a such a way that each grid cell contains $O(n^\epsilon)$ sub-cells of the next level, bottom level cells contain at most $O(n^\epsilon)$ points and the data-structure consists of $O(\epsilon^{-1})$ levels. Intuitively, the algorithm aims to match every point within the smallest grid cell possible, however, certain cells might not have the same number of red and blue points.

A crucial difference between our implementation and that of \cite{AV04} is that we describe a process which constructs the matching in a bottom-up manner, starting with the smallest cells and progressing towards the cell containing all of the input, whereas the prior work iterates in a top-to-bottom order. This change allows for a shorter algorithm description; however, it requires a different but arguably simpler analysis, which we present in Appendix~\ref{sec:app:static-proof}. 

As bottom cells contain only a small number of input points, they may select a maximal color-balanced subset of them, match its points optimally using a polynomial running time static algorithm, and forward the remaining points to their parent cell.

If a set of points can't be matched within a cell due to color dis-balance, we can argue that an optimal cost matching must match the dis-balance with points outside of the cell, and hence with longer edges. This allows us to introduce some slack and represent these forwarded points implicitly by the center of their respective cell and their cardinality. 

Larger cells aiming to match these forwarded implicitly represented points can efficiently find an implicit matching of the underlying points by an optimal polynomial running time geometric transportation algorithm. These implicit matchings can then efficiently be turned into an actual matching of the underlying points of roughly the same cost.

To dynamize the algorithm, we show that, due to an update, the matching structure might only change in cells containing the updated points. After an update for each affected cell in a bottom-up manner we carefully adjust the matchings calculated. Since in each cell we either have $O(n^\epsilon)$ points or a larger set of points represented with $O(n^\epsilon)$ sub-cell center points, this can be done with polynomial running time static sub-routines for each affected cell. As the nested grid contains $O(\epsilon^{-1})$ levels, the update time works out to $O(n^{\epsilon} \cdot \epsilon^{-1})$.

Updating the matching of a bottom cell is straightforward, as we may run an optimal polynomial time algorithm on their updated points. However, the sub-cell center-points of intermediate cells may represent $\Omega(n)$ input points implicitly. Hence, if the implicit matching of the cell significantly changes due to an update, it may take $\Omega(n)$ time to update the actual output matching within the cell.

To overcome this difficulty, we first carefully analyze the update process and show that the points matched in each cell may only change by $2$ due to an update. Afterwards, we present a graph shortest $s-t$ path sub-routine based procedure which calculates an updated version of the geometric transport solution stored in the cell. This solution only differs in at most $O(n^\epsilon)$ edges compared to its unupdated state. In turn, this allows us to efficiently update the matching of the actual input points implicitly represented by the geometric transportation sub-solutions.

\subsection{Related Work}\label{sec:related}
Numerous results exist on special cases and generalizations of Euclidean bi-chromatic bipartite matching across various computational settings. In the static setting, an exact solution can be found through minimum cost max-flow graph algorithms in $\hat{O}(n^2)$ time \cite{focs/BrandLNPSS0W20}. For restricted cases of the problem, \cite{compgeom/Sharathkumar13} has presented an algorithm with $\hat{O}(n^{3/2})$ running time. In the approximate setting, \cite{stoc/AgarwalCRX22} presented the first deterministic $\tilde{O}(n)$ time algorithm for arbitrarily tight approximations. Due to space restrictions, a more detailed overview and discussion on related work can be found in Appendix~\ref{app:sec:related-works}.

\paragraph{Paper Outline:} 
We describe a static algorithm and its dynamic implementation in sections \ref{static} and \ref{dynamic} respectively. 
In Section~\ref{sec:experiments}, we demonstrate the practicality and efficiency of the dynamic algorithm for estimating the $1$-Wasserstein distance.

\section{Preliminaries}
\label{sec:prelim}


\paragraph{The Euclidean matching problem} 
Given a set of red points \(A\) and a set of blue points $B$ in the plane, where  \(|A| = |B|=n\), a \textit{matching} is a bijection of the form \(\mu: A\rightarrow B\). The \textit{cost} of a matching $\mu$ is given by \(c(\mu) = \sum_{a\in A} \|a-\mu(a)\|_2\).
The objective is to find a matching of \emph{minimum} cost. 
We say that a pair of points \(e := (a, b)\), referred to as an \emph{edge}, belongs to matching \(\mu\) if \(b=\mu(a)\). 
We define a matching $\mu$ to be $\alpha$-approximate for $\alpha \geq 1$ if $c(\mu) \cdot \alpha \leq c(\mu^*)$ for any optimal solution $\mu^*$.

To simplify presentation, we assume that $X=A \cup B \subseteq D\times D$ for some $D = \operatorname{poly}(n)$, where $D$ is a power of $2$. Throughout this paper we make the standard assumption that the \emph{spread} of the input points, i.e. the ratio between the largest and the smallest inter-point distance in \(X\),  remains bounded by some large $U  = \operatorname{poly}(n)$.

A generalization of the matching problem, supporting point multiplicities, is the so-called \emph{Euclidean transportation problem}, defined as follows.
\paragraph{The Euclidean transportation problem} Let \(A, B\subset \mathbb{R}^2\) be any two sets of points such that each point \(a\in A\) has a non-negative integer \emph{supply} \(s_a\geq 0\) and each \(b\in B\) has a non-positive integer \textit{demand} \(d_b\leq 0\) satisfying $\sum_{a\in A} s_a  + \sum_{b\in B} d_b = 0.$ 
An \textit{assignment}, \(\gamma: A \times B \rightarrow \mathbb{Z}_{\geq 0}\), is a non-negative set of weights on the edges such that $ s_a  = \sum_{b\in B} \gamma (a, b),  \forall a \in A$ and $-d_b  = \sum_{a\in A} \gamma (a, b), \forall b \in B $.

 The \emph{cost} of an assignment $\gamma$ is \(c(\gamma)= \sum_{a\in A,b\in B} \gamma(a, b)\|a-b\|_2\). An \textit{optimal} assignment for the Euclidean transportation problem is an assignment 
 of \emph{minimum} cost.

\begin{theorem}
\label{thm:transport}
\cite{AtV95} There exists a deterministic algorithm which, given an instance of the Euclidean transportation problem $A,B \subseteq \mathbb{R}^2$, $s \in \mathbb{Z}_{\geq 0}^A$, $d \in \mathbb{Z}_{\leq 0}^B$, returns an optimal assignment  {$\gamma : A \times B \rightarrow \mathbb{Z}_{\geq 0}$} in time $O(n^{5/2} \log(n \log M))$, where $M = \max( \left \lVert s \right \rVert_{\infty}, \left \lVert d \right \rVert_\infty)$, $n = |A \cup B|$.
\end{theorem}

The results of this paper rely on a standard data structure for representing points in a geometric setting (see \cite{AV04,stoc/AgarwalCRX22}), the underlying and restricted \(p\)-trees, two quadtree-like structures. Throughout the paper think of the parameter $p$ as some $n^\epsilon$ sized power of $2$.

 \paragraph{The underlying and the restricted $p$-tree} 
We define a \emph{cell} of side-length $L>0$ as a square of the form $[a, a+L) \times[b,b + L)$ for some $a,b\in \mathbb{R}$.
We will frequently subdivide cells of side-length $L$ into \emph{subcells} of side length $L/p$.
We will call this a \emph{grid} of side length $L/p$.


For an integer \(p\geq 2\), which is also a power of $2$, and a point set \(X\), we define an \textit{underlying} \(p\)-\textit{tree} \(T\) on \(X\) to be a quadtree-like structure with larger branching factor (see e.g., \cite{AV04}), formally described as follows.

\begin{enumerate}[noitemsep,topsep=0pt,parsep=2pt,partopsep=0pt]
    \item[(1)] The \textbf{root} $\bar{r}$ consists of a bounding cell with side-length $D$, containing all points in \(X\), and its subdivision into \(p^2\) smaller subcells. The root $\bar{r}$ has \(p^2\) children, denoted by $\mathbb{C}_{\bar{r}}$, corresponding to its subcells, each of side-length \(D/ p\). 
    The root is at level $0$ of \(T\).
    \item[(2)] Each \textbf{internal node} $v$ at level \(i>0\) corresponds to one of the subcells of its parent node. 
    Thus, an internal node $v$ at level \(i\) consists of a cell of side length \(D/p^i\), and its subdivision into \(p^2\) smaller subcells, each of size \(D/ p^{i+1}\times D/ p^{i+1}\). 
\end{enumerate}

For a node $v$ of the underlying $p$-tree $T$, let $X_v$ be the set of points from $X$ belonging to the bounding cell of $v$. We say that a point \(x\in X\) \textit{belongs to node} \(v\) if \(x\in X_v\). 
Now, notice that for some nodes \(v\) (on both low and high levels) the set \(X_v\) might be empty. To avoid storing unnecessary empty cells (which would result in a slower data structure), we define the \textit{restricted} \(p\)-\textit{tree} or shortly \(p\)-\textit{tree}, which is a subtree of the underlying \(p\)-tree, where the root and leaves are defined as follows:

\begin{enumerate}[noitemsep,topsep=0pt,parsep=4pt,partopsep=0pt]
    \item[(1)] The \textbf{root} \(r\) is a node of the underlying \(p\)-tree such that \(X_r=X\), but which does not have a child \(v\) with \(X_v=X\). In other words, \(r\) is the node with \(X_r=X\) located on the lowest possible level of the underlying \(p\)-tree.
    \item[(2)] A \textbf{leaf} of the \(p\)-tree is a cell \(v\) such that \(|X_v|\leq p^2\), but whose parent 
    \(u\) has \(|X_u|>p^2\).
\end{enumerate}
%

As we insert or delete points in the dynamic setting, notice that the restricted \(p\)-tree might change over time: the root, as well as the leaves might change to higher or lower levels of the underlying \(p\)-tree. However, as we assume that the spread of the pointset \(X\) is bounded by \(U\) at all times, we know that the height of the restricted \(p\)-tree is bounded by \(d:=\log_p U\) and is therefore constant if \(p\) is chosen to be sublinear in \(n\).

It can be shown that a restricted \(p\)-tree w.r.t. set \(X\), \(|X|=n\), can be maintained under insertions and deletions of points in \(O(p^2+\log_p U)\) amortized update time (see e.g. \cite{de2007efficient}).

\paragraph{$z$-ordering} The \(z\)-ordering of the point-set \(X\) w.r.t. an underlying \(p\)-tree \(T\) is defined via specific ordering of children nodes in \(T\). Starting from the root of \(T\), we traverse the children cells of each internal node row by row, then recursively traverse the subcells of each cell until a subcell contains only one point.

\section{Basic Framework and the Static Algorithm}        \label{static}

The starting point of our work is the static algorithm for computing bi-chromatic Euclidean matchings due to \cite{AV04}. 
Specifically, their top-down algorithm computes an $O(1/\varepsilon)$-approximate matching for an instance $(A,B)$ in time $O(n^{1+\varepsilon})$ for sufficiently small constants $\varepsilon>0$. 
We begin by presenting our modified variant of this algorithm for the static setting, referred to as $\staticmatch(A,B)$. 
The next section shows that our variant can be extended to a bottom-up algorithm for the fully dynamic setting achieving sub-linear update time.

For the input sets \(A, B\) with bounded spread $U$ we first apply a random shift to all the points by a fixed random vector $r \in [U]^2$ to get $A' = \{a+r:a\in A\}$ and  ${B' = \{b+r: b\in B\} }$. 
From now on, we are only interested in sets \(A'\) and \(B'\). 
Slightly abusing notation, we will simply refer to these sets as \(A\) and \(B\).
Our first step is to construct a restricted $p$-tree \(T\) on the point set \(X:=A\cup B\). 
Using \(T\) in a 
\emph{bottom-up} manner, we obtain a matching between the point set $A$ and $B$ as follows. 
Intuitively, as the number of unmatched red and blue points belonging to some node might not be balanced, we greedily match as many red-blue pairs of points as possible at that node.
Handling the unmatched points is deferred to the parent node, which matches as many excess points from its children as possible, and so forth. 
Therefore, for any node \(v\) of \(T\), we keep track of the set of red points \(Y_v(A)\) and blue points \(Y_v(B)\) that we match at that node, and the set of unmatched monochromatic points that are in excess, denoted by \(E_v\).


An essential element of our algorithm is the choice of the matching algorithm used throughout the traversal of \(T\).
Observe that the excess points forwarded to an internal node of $T$ by its children could be of size $\Omega(n)$. To overcome this problem, our algorithm will rely on a data structure representing excess points forwarded by node $v$ of $T$ by the center point of the cell of $v$. Using this data structure we are able to efficiently compute (and maintain) an \textit{implicit matching} of the unmatched points at all $v \in V$, which we can efficiently turn into an $\textit{explicit matching}$ of the underlying points when queried.

\subsection{Sub-instances via transportation solver} 
\label{sec:transportation}

An \textit{implicit matching} is an optimal assignment $\gamma_v^*[Y_v] : v \in T$ associated with a vertex $v$ of $T$ of an Euclidean transportation problem defined by some inputs $(Y_v(A), Y_v(B), s, d)$.

To make sense of the above vague definition, we describe procedures \textsc{Implicit-Matching} and \textsc{Explicit-Matching} that will act as a sub-routines in both static and dynamic algorithm for Euclidean matchings. The purpose of these functions is to obtain two representations of a maximal matching between specific points forwarded to an internal vertex from its children in a restricted \(p\)-tree.

Given a subset of points belonging to a node in the $p$-tree, in the \textsc{Implicit-Matching}, we first define the notion of a \emph{aggregated} sub-instance for that node. To aggregate the points of a given cell, we move all of them to the cell's center.
We define an Euclidean transportation problem on this aggregated sub-instance, and the resulting optimal assignment will correspond to an \emph{implicit matching} of $v$. 
\textsc{Explicit-Matching} converts this implicit matching into a matching of the original subset of points, which we refer to as the \textit{explicit matching}.


\paragraph{$\textsc{Implicit-Matching}$ procedure}

The input of the procedure is a vertex \(v\) of \(p\)-tree \(T\), the monochromatic sets of points \(\{E_u:u\in \mathbb{C}_v\}\) (denote their union by $\mathcal{E}$, and partition into blue $\mathcal{E}_B$ and red $\mathcal{E}_A$ points) forwarded to \(v\) from its children and their cardinalities \(\{|E_u|:u\in \mathbb{C}_v\}\). Without loss of generality, assume that $|\mathcal{E}_B| \geq k = |\mathcal{E}_A|$.

Its output consists of an optimal assignment of a transportation problem (i.e., an implicit matching) defined by $k$ points of $\mathcal{E}_B$ and all points of $\mathcal{E}_A$ (i.e., the points we can match within the cell of $v$) and $|\mathcal{E}_B| - k$ blue points, which will be forwarded to the parent of $v$ in $T$.

\textit{Choosing the points to match.} Let $Y_v(B)$ stand for the first $k$ points of $\mathcal{E}_B$ with respect to some ordering (say the z-ordering). Let $E_v = \mathcal{E}_B \setminus Y_v(B)$ be the \textit{excess set}, i.e. the set of points not matched at $v$, $Y_v(A) = \mathcal{E}_A$ and $Y_v = Y_v(A) \cup Y_v(B)$. Excess set $E_v$ will be forwarded to the parent of $v$.

\textit{Computing the implicit matching.}
We define an instance of Euclidean transportation problem over the center points of all $p^2$ sub-cells of the cell of $v$. Note that every such sub-cell $u \in \mathbb{C}_{v}$ contains a monochromatic subset of $\mathcal{E}$. Define $A^c$ and $B^c$ to be center points of sub-cells containing only red and blue points of $Y_v$ respectively. We set $s_{a'}$ for $ a' \in A^c$ and $-d_{b'} $ for $ b ' \in B^c$ to correspond to the number of red and blue points of $Y_v$ of the cells with centers $a'$ and $b'$, respectively.

The optimal assignment $\gamma^*_v[Y_v]$ obtained by solving the Euclidean transport problem defined by $(A^c,B^c,s,d)$ using the algorithm of Theorem~\ref{thm:transport} will be returned as the implicit matching of vertex $v$. Crucially, observe that setting up the Eucledian transport instance doesn't require the knowledge of the exact location of vertices in $Y_v(A),Y_v(B)$ just their cardinalities in the sub-cells of $v$.

\paragraph{\textsc{Explicit-Matching} procedure} The inputs of the procedure are sets \(Y_v(A)\) and \(Y_v(B)\), as well as implicit matching \(\gamma^*_v[Y_v]\) between those points at an internal node \(v\). Its output is a matching between \(Y_v(A)\) and \(Y_v(B)\).

\begin{figure*}
    \centering
    
        \includegraphics[width=0.20\linewidth]{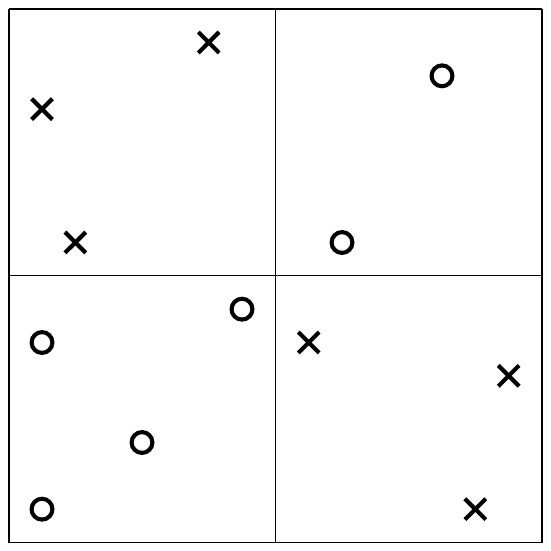}
    \hspace{0.5cm}
        \includegraphics[width=0.20\linewidth]{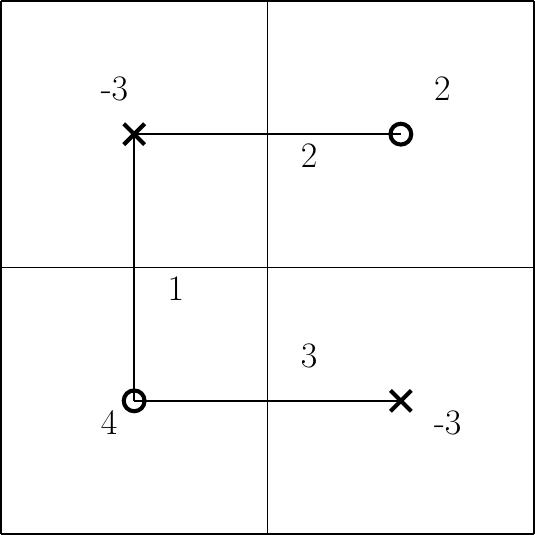}
    \hspace{0.5cm}
        \includegraphics[width=0.20\linewidth]{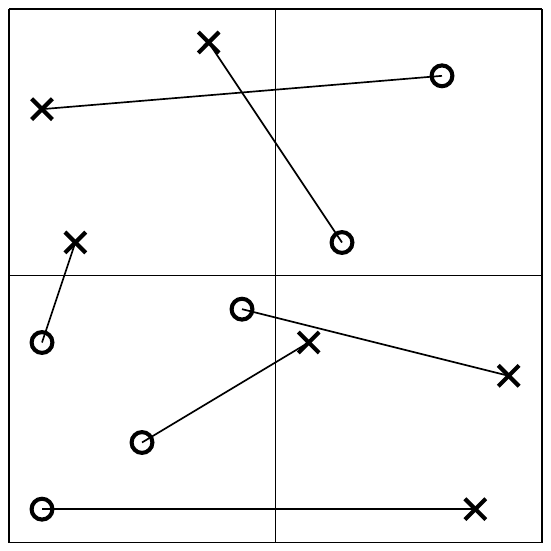}
    \caption{
        Examples of an instance (left), implicit matching of the aggregated sub-instance (middle), and the resulting matching sub-instance (right). 
        In the left picture, the red points are represented by circles, while the blue points are represented by crosses. 
        The middle picture corresponds to the implicit matching on the aggregated sub-instance, with numbers next to vertices representing their demands and supplies, while numbers next to edges represent the weighting of the edges. The right picture shows the corresponding matching of the input points.
    }
    \label{fig:moved_match}
\end{figure*}

The procedure naturally translates the assignment \(\gamma^*_v[Y_v]\) to a perfect matching \(\mu'_v[Y_v]\) between the points \(Y_v(A)\) and \(Y_v(B)\) as follows: match $\gamma^*_v[Y_v](a',b')$ many different pairs of points of the form $(a,b)$, where $a$ and $b$ lies in the sub-cell whose centers are $a'$ and $b'$ respectively.
We refer to the latter matching as an \emph{explicit} matching of node $v$ w.r.t. its corresponding point set $Y_v$. For an example involving both implicit and aggregated sub-instances, see Figure \ref{fig:moved_match}. 

In the following lemmas, we summarize the above algorithms and their running times.

\begin{lemma} \label{lm:implicitMatching}
    Let \(v\) be an internal node of a \(p\)-tree \(T\) on $X := A \cup B $ with \(|A|=|B|=n\). Let \(\mathcal{E}=\{E_u:u\in \mathbb{C}_v\}\) be the pointers to some monochromatic sets of $v$'s  children forwarded to \(v\), as well as their cardinalities \(\{|E_u|:u\in \mathbb{C}_v\}\). 
    There is an algorithm \textup{$\textsc{Implicit-Matching}(v,\mathcal{E}, \{|E_u|:u\in \mathbb{C}_v\})$} that chooses balanced red and blue sets  \(Y_v(A)\) and \(Y_v(B)\), computes an implicit (perfect) matching \(\gamma^*_v[Y_v]\) between them, in time 
    \(O(p^5 \log p \log n)\). Additionally, the function returns the (pointer to the) monochromatic set of unmatched points \(E_v\) at node \(v\).
\end{lemma}

A crucial property of the Lemma~\ref{lm:implicitMatching} is that it has a running time only polylogarithmic with respect to $n$ although $\mathcal{E}$ could consist of $\Omega(n)$ points. This can be naturally achieved through an implicit representation of the point sets \(\{E_u:u\in \mathbb{C}_v\}\). We formalize this process in Appendix~\ref{sec:app:implicit-explicit}.

Now, as described above, it is easy to convert an implicit matching to a matching of the input points.

\begin{lemma}
\label{lemma:explicit}
Let $T$ be a $p$-tree 
on $X = A \cup B$ and $d$ its depth. The following statements hold:
\begin{enumerate}[noitemsep,topsep=0pt,parsep=2pt,partopsep=0pt]
    \item Given a vertex \(v\) of a \(p\)-tree \(T\), and an implicit matching \(\gamma^*_v[Y_v]\) w.r.t. \(v\) and set \(Y_v\), 
    the corresponding explicit matching \(\gamma'_v[Y_v]\) can be obtained in time \(O(|Y_v|d)\).
    \item Given tree \(T\) where each leaf stores a matching and each non-leaf node an implicit matching, the corresponding perfect matching on $X$ can be reported in \(O(n)\) time.
\end{enumerate}
     
\end{lemma}

\subsection{Static Algorithm}

Next, we present the formal description of the bottom-up algorithm. 
To this end, we define the subroutine $\match (v)$ as follows. 
If node $v$ is \emph{leaf} of the restricted \(p\)-tree: 
\begin{enumerate}[noitemsep,topsep=0pt,parsep=2pt,partopsep=0pt]
        \item  {\bf Determine sets $Y_v(A), Y_v(B)$ and $E_v$:}  
        Let 
        $Y'_v(A)=X_v\cap A$ be the red points, 
        $Y'_v(B)=X_v\cap B$ the blue points, and 
        $\eta_v = |Y'_v(A)|-|Y'_v(B)|$ the difference in cardinality, 
        at node \(v\). 
        
        If \(\eta_v = 0\), we set \(E_v = \emptyset\). 
        If $\eta_v>0$, 
        we assign a subset of $\eta_v$ red points chosen as the last \(\eta_v\) points in the z-ordering from \(Y'_v(A)\) to $E_v$, and set \(Y_v(A)=Y'_v(A)\setminus E_v\) and \(Y_v(B)=Y'_v(B)\) (in the same way as presented in Section \ref{sec:transportation}). 
        Analogously, if $\eta_v<0$, we assign a subset of $\eta_v$ blue points from \(Y'_v(A)\) to $E_v$, and set \(Y_v(B)=Y'_v(B)\setminus E_v\) and \(Y_v(A)=Y'_v(A)\). 
        We refer to $E_v$ as the \textit{excess points} at \(v\). 
        \item {\bf Compute matching:}  Obtain a matching \(\mu_v\) between red points \(Y_v(A)\) and blue points \(Y_v(B)\) using any efficient exact algorithm (e.g. the Hungarian algorithm \cite{munkres}), and compute its cost \(c(\mu_v)\). 
        \item  Return the cost and the excess set \((c(\mu_v), E_v)\). \\ 
    \end{enumerate}

\noindent Otherwise, node $v$ is \emph{internal}: 

\begin{enumerate}[noitemsep,topsep=0pt,parsep=2pt,partopsep=0pt]
    \item 
         \label{implicit_step} {\bf Compute implicit matching:} 
        For each child \(u\) of $v$, let \(E_u\) be the set of excess points returned by \(\textsc{Match}(u)\). 
        Invoke algorithm \textsc{Implicit-Matching}$(v,  \mathcal{E}, \{|E_u|:u\in \mathbb{C}_v\})$ from Section~ \ref{sec:prelim} (see Lemma~\ref{lm:implicitMatching}). Let $\gamma^*_v[Y_v]$ be the resulting implicit matching w.r.t. node \(v\) and set $Y_v$ computed via the algorithm, and $E_v$ the set returned by the algorithm.
        Set the estimate of the cost 
        for the matching in $v$ to \(c(\mu_v) := c(\gamma^*_v[Y_v])+ \sum_{u\in \mathbb{C}_v} c(\mu_{u})\). 
        \item Return the cost and excess set \((c(\mu_v), E_v)\). \\

    \end{enumerate}

Our algorithm \staticmatch$(A, B)$ first invokes the $\match(r)$ subroutine, where \(r\) is the root of \(T\) on \(A\cup B\). 
Note that this yields implicit matchings \(\gamma^*_v[Y_v]\) at internal nodes \(v\) of \(T\). 
Using Lemma \ref{lemma:explicit}, Part~(2), we can transform the implicit matchings \(\gamma^*_v[Y_v]\) of the aggregated versions of $Y_v$ into matchings of the original points \(\mu'[Y_v]\) of roughly the same cost (see Claim~\ref{cl:static-proof:mamdash} for a formal bound on the slack introduced by the aggregation).
Finally, we return a maximal matching \(\mu_r\) on the entire point set \(X\), along with its already computed cost \(c(\mu_r)\).

Our analysis of the algorithm builds on the following key idea from \cite{AV04}: when all points are shifted randomly by the vector $r$, the probability of an edge being cut by a grid in a node of the $p$-tree is proportional to the length of the edge and inversely proportional to the spacing of the grid.

However, the algorithm of \cite{AV04} works in a top-to-bottom approach (building the matching starting from the root and progressing to the leaves) in contrast to our algorithm, which takes a bottom-up approach. The bottom-up approach allows us to guarantee that for any node $v \in T$ the number of nodes not matched within the cell of $v$ simply corresponds to the difference in the number of blue and red nodes inside the cell of $v$. Intuitively, this implies that points are matched to near-by neighbor whenever possible.

This results in an arguably simpler analysis differing from that of \cite{AV04}, and crucially relying on the concept of implicit matchings. We provide a self-contained analysis of the static algorithm in Appendix~\ref{sec:app:static-proof}.

\begin{restatable}[Static]{theorem}{static}
\label{thm:static}
    For any Euclidean matching instance $(A,B)$,  with $|A| = |B|=n$, and spread $U \leq n^c$, for some constant \(c > 0\), and for any constant $\epsilon \in (0,\frac{1}{10})$, 
    algorithm \textup{\textsc{Static-Matching}$(A,B)$} computes an  $O(1/\epsilon)$-approximate Euclidean matching with high probability in $ O( n^{1+\epsilon} \cdot \epsilon^{-1})$ time. 
\end{restatable}

\section{Dynamic Algorithm}\label{dynamic}


In this paper, we present two dynamic data structures for maintaining perfect matchings under insertions and deletions of pairs of red and blue points. 
Both data structures naturally convert our static, bottom-up algorithm from the previous section into a dynamic one.

Our first algorithm \(\textsc{EuclMatch1}\) implicitly maintains a matching for each node of the tree. 
This allows us to (i) estimate the cost of the entire matching after each update and (ii) report the current approximate matching upon a query in time proportional to its size. 
Our second algorithm \(\textsc{EuclMatch2}\) explicitly maintains the output matching at all times. We defer the second algorithm to Appendix~\ref{sec:app:advanced} due to page limitations.

\paragraph*{Initialization}
For any given set of points \(A\) and \(B\), we will next describe how to initialize a matching by essentially running the bottom-up algorithm $\staticmatch$ on the instance $(A,B)$. Formally, the $\textsc{Initialize}(A,B)$ initializes the restricted \(p\)-tree \(T\) w.r.t. \(X:=A\cup B\), and the following corresponding values: the set of unmatched points $Y'_v(A), Y'_v(B)$, the excess set \(E_v\) at each node \(v\) of \(T\), the implicit matching $\gamma^*[Y_v]$ and the cost \(c(\mu_v)\) of the matching w.r.t. the subtree rooted at each internal node, and the matching \(\mu_v\) and its cost \(c(\mu_v)\) at each leaf. This process takes $O(n^{1+\epsilon})$ time by Theorem~\ref{thm:static}.






\paragraph*{Handling updates}

As the procedures for insertions and deletions are similar (see Appendix \ref{app:basic}), we focus on presenting the insertion procedure. Without loss of generality assume that we insert a red point \(a\) to \(A\). The algorithm \(\textsc{Insert}(a)\) (see Algorithm~\ref{alg:insert}) proceeds in two steps: updating the restricted \(p\)-tree \(T\), and updating the matchings at each \textit{affected} node, i.e. node \(v\) where \(a\in X_v\). 

First, updating \(T\) follows a standard approach, similar to the method described in e.g. \cite{har2011geometric}. Namely, one or both of the following cases can happen:
\begin{enumerate}[noitemsep,topsep=0pt,parsep=2pt,partopsep=0pt]
    \item After insertion of \(a\) to \(A\), a leaf \(v\) of \(T\) might contain more than \(p^2\) points. In this case, \(v\) now becomes an internal node of \(T\), while a new subtree has to be built under \(v\). We refer to these nodes as \textit{new subroots}, while all other nodes belonging to this subtree are called \textit{marginal nodes}. Finally, all nodes in \(T\) that are not marginal are refered to as \textit{non-marginal nodes}.
    \item If \(a\) does not belong to the current root of the restricted \(p\)-tree, a internal node of the underlying \(p\)-tree that is on a higher level than the previous root now becomes the new root of \(T\).
    
\end{enumerate}

Second, we define the recursive procedure \(\textsc{Insert-Update}(v, a)\) which updates the matchings at each affected node of \(T\) in a bottom-up manner (see Alorithm \ref{alg:insert-update}).
Specifically, we define \(\textsc{Insert-Update}(v, a)\) for three types of affected nodes of \(T\): non-marginal internal nodes, non-marginal leaves, and new subroots. We focus on presenting the case for non-marginal internal nodes, as the other two cases are simpler, and can be found in Appendix \ref{app:basic}.

If a node \(v\) is a non-marginal internal node, the procedure \(\textsc{Insert-Update}(v, a)\) first updates set \(E_u\) for the affected child \(u\) of \(v\), via \(\textsc{Insert-Update}(u, a)\). Then, we update the excess set \(E_v\) accordingly. Finally, we invoke the \(\textsc{Implicit-Matching}(\cdot)\) procedure to obtain an implicit matching for the remaining points \(Y_v\) at \(v\).

Crucially, observe that throughout this process the affected nodes of $T$, that is nodes at which either the implicit matching or the excess set changes, lie along a path from a leaf to the root of $T$. Hence, the most costly operation of the algorithm, the updating of the implicit matchings, only has to be repeated at a small number of nodes.

The correctness of the algorithm follows from the algorithm always satisfying that after handling an update its output corresponds to the run of a static algorithm on the current state of the input, up until the different choices the underlying optimal bi-chromatic matching algorithm (Hungarian algorithm) and optimal geometric transport algorithm (see Theorem~\ref{thm:transport}) makes. Note that the choice of underlying exact algorithms is arbitrary as long as they run in polynomial time.






\paragraph*{Queries} 
To report the changes to the matching, the algorithm converts the implicit matchings at the affected nodes to an explicit matching of \(X\) via Lemma \ref{lemma:explicit}. To query the cost of the approximate solution the algorithm simply returns the approximate cost value 
\(c(\mu_r)\) that is maintained at root \(r\) of \(T\).

The following theorem summarizes the results of our simpler dynamic algorithm.

\begin{theorem}
\label{thm:basic}
    Let $\eps \in (0,1]$ and \((A, B)\) two point sets in \(\mathbb{R}^2\) with \(|A|=|B|=n\), and spread $U< n^c$ for some constant \(c > 0\). 
    There is a dynamic data structure that supports 
    insertions and deletions in \(O(n^{\varepsilon} \cdot \varepsilon^{-1})\) \textit{worst-case} update time and maintains an expected $O(1/\epsilon)$-approximate matching between the dynamic set $(A,B)$. 
    Initialization takes \(O( n^{1+\varepsilon} )\) time.
    Reporting the matching, the changes to the matching when an insertion or deletion occurs (recourse), and the cost of the matching, take $O(n)$, $O(n)$ and $O(1)$ time, respectively. 
\end{theorem}

\paragraph{Improved dynamic algorithm}

Ideally, the algorithm would maintain the solution matching explicitly. This does require however, that the explicit matchings implicitly represented at all nodes of $T$ are 
explicitly maintained at all times. This requires a careful propagation of changes throughout the $p$-tree and the ability to update explicit matching efficiently when the underlying implicit matching changes. To dynamize the maintenance of explicit matchings, we rely on the fact that a single insertion might only change the excess set stored at each node by  $1$ point.

This implies that the difference between the updated and pre-updated matching lies along a short alternating path of old and new matching edges. We find this augmenting path using a graph shortest-path based sub-routine, which, due to page limitations, we defer to Appendix~\ref{sec:app:advanced}.

\section{Experimental Evaluation}
\label{sec:experiments}

With this work, we provide\footnote{\url{https://github.com} Anonymized for submission, see supplemental file.} an opensource, single-threaded C++ implementation of our algorithm in Appendix~\ref{app:update}.
To demonstrate the practicality and effectiveness our proposed algorithms, we conducted experiments on a $2.2$~GHz \verb|Ubuntu 22.04.4| system with AMD~Opteron~6174, using synthetic and real world data sets. 
For reproducibility and comparability, we report on measurements for the same 2D distributions and data sources used in the recent work~\cite{GattaniRS23}. 
For the {\em synthetic data}, we drew the point coordinates from the uniform distribution on integers between $1$ and $500$ to obtain the \verb|Uniform| datasets.
For the \verb|Gaussian| datasets, we drew the point coordinates from the normal distribution ($\mu=0.5, \sigma=0.25$) prior to scaling by a factor of $500$ and rounding to integers.
For the {\em real data}, we extracted all trips from the Yellow-Cap \verb|Taxi| dataset\footnote{See \url{https://www.nyc.gov/site/tlc/about/tlc-trip-record-data.page}.} of December 2009, and, following~\cite{GattaniRS23}, extracted those having at least $3$~minutes duration, at most $110$~mph speed, and latitude/longitude in the range of $[-74.5,-73.5]\times[40,41]$,
yielding $14\,324\,017$~trips with one `Pickup' and `Dropoff' location each.



\subsection{Experimental Results}

\paragraph{Speedup: Dynamic vs Static Approximation Algorithms}
To measure the speedup or of dynamic 
over baseline approximations, we measured the update time and time to compute a approximate matching from scratch, for various levels of approximation quality.
Figure~\ref{fig:SpeedUp} shows the speedup factors ($p=8$) for inserting $\leq\!8\,000$ samples from  \verb-Uniform- vs \verb-Guassian- (top) and \verb-Pickup- vs \verb-Dropoff- (bottom). 

\paragraph{Insertion-Workloads: Estimating Wasserstein distance using Exact vs Approximate Matchings}
To demonstrate the usability of approximations of minimum cost bi-chromatic matchings to estimate the Wasserstein distance, we estimated the empirical Wasserstein distance using exact and approximate matchings on identical and different distributions for various levels of approximations and sample sizes.
Figure~\ref{fig:wasser-exact-vs-apx} shows the empirical $1$-Wasserstein distance of two \verb-Uniform- distributions (top) using exact minimum cost matchings and approximations with $p=2$, $p=8$, and $p=32$, for up to $2\,000$ samples.
The bottom figure shows the same estimates for \verb-Uniform- vs \verb-Gaussian-.

\paragraph{Insertion-Workloads: Convergence and Update Time for Very Large Sample Sizes}
To demonstrate the scalability of our approach, we measured the convergence and update time of our dynamic approximations for sample sizes up to $1\,000\,000$, rendering the exact solvers impractical.
Figure~\ref{fig:large} (top) shows the convergence of empirical $1$-Wasserstein distance, using $p=2,~4,~8$ for approximations, for the \verb|Uniform| vs \verb|Uniform| benchmark (bottom three curves) and the \verb|Uniform| vs \verb|Gaussian| benchmark (top three curves). 
The bottom figure shows the update time, in \emph{milliseconds}, of our dynamic algorithm on both benchmark data sets.

\paragraph{Fully-Dynamic Workloads: Drift of Pickup vs Dropoff distributions over time}
To demonstrate the effectiveness of our dynamic algorithms to monitor the drift of time-dependent, spatial distributions, we sorted the trips in the \verb|Taxi| dataset by pickup time and use a sliding window, having a width of $10\,000$ samples, to obtain update sequences that contain insertions and deletions.
Figure~\ref{fig:drift-real} (top) shows the empirical $1$-Wasserstein distance between the Pickup and Dropoff locations, using approximations with $p=4$ and $p=16$.
The bottom figure shows the update time of our fully dynamic algorithm on the \verb|Taxi| benchmark.


\subsection{Discussion of experimental results}
The experimental results show that the dynamic algorithm is orders of magnitudes faster than computing static approximations for minimum cost bi-chromatic matching (Figure~\ref{fig:SpeedUp}).
In practice, the approximate matching costs are within a very small factor (i.e. $<2$) of the minimum bi-chromatic matching cost, which enables very accurate and effectiv estimation the $1$-Wasserstein distance (Figure~\ref{fig:wasser-exact-vs-apx} and top part of Figure~\ref{fig:large}).
As suggested by our Theorem~\ref{main}, we observe an tradeoff between approximation quality and update time (bottom part of Figure~\ref{fig:large}). Updates of the dynamic algorithm took typically between one and ten millisecond and showed a clear separation, between inputs with zero and non-zero distance, at sample sizes around $10\,000$ samples.
The time dependent spatial distributions of Pickup and Dropoff locations in the \verb|Taxi| dataset show a drift in the empirical $1$-Wasserstein distance (Figure~\ref{fig:drift-real} top), and the dynamic algorithm allows for effective monitoring of the change even on large window sizes on real data sets (Figure~\ref{fig:drift-real} bottom).



\printbibliography


\onecolumn

\appendix

\ifdefined\todos
\fi

\section{Related Works}

\label{app:sec:related-works}

{\em Static Algorithms:} 
The classical Hungarian algorithm~\cite{kuhn1955hungarian} allows computing an optimal Euclidean bi-chromatic matching in $O(n^3)$ time.
\cite{AtV95} presented a Primal-Dual based scaling algorithm for points from $\mathbb{R}^2$ that, supporting points with multiplicity at most $W$, runs in $O(n^{5/2}\log(n)\log(W))$ time.
Near-quadratic algorithms for computing an optimal matching (of points with multiplicity) can be derived from recent advances for minimum cost max-flows in graphs~\cite{focs/BrandLNPSS0W20}. 
%

Sub-quadratic algorithms are known for the following two important cases.
In case $d=2$ and all coordinates are integers $\leq\!U$, \cite{compgeom/Sharathkumar13} presented an $O(n^{3/2+\delta}\log(nU))$, for any fixed $\delta\in (0,1]$.
In case both $A,~B$ are i.i.d. samples (from two distributions), 
\cite{GattaniRS23} recently presented an 
$\widetilde O(n^{2-\frac{1}{2d}} \log (U) \Phi(n))$, where $\Phi(n) = \operatorname{poly}(\log n)$ for $d=2$, 
expected time algorithm, and showed its practicality for estimating the $1$-Wasserstein distance of distributions on $\mathbb{R}^2$ for $n\leq50\,000$ samples in approximately $700$ seconds.

{\em Static Approximations:} 
 \cite{AV04} presented a top-down algorithm that computes, with with high probability, an $O(\log \frac{1}{\eps})$-approximation in $O(n^{1+\eps})$-time, for any fixed $\eps \in (0,1]$. \cite{stoc/SharathkumarA12}, \cite{jacm/RaghvendraA20} obtained an $O(n \operatorname{poly}(\log n,\eps^{-1}))$ time algorithm that, using augmenting paths in a randomly shifted quad-tree, computes with high probability a $(1+\eps)$-approximate matching.
The $\operatorname{poly}(\log n, \eps^{-1})$-factor was further improved in~\cite{swat/AgarwalRSS22}.
Recently, \cite{stoc/AgarwalCRX22} presented the first deterministic $(1+\eps)$-approximation algorithm with near-linear $n(\frac{\log n}{\eps})^{O(d)}$ time. 

{\em Variants:}
Using spanners and fast minimum cost max-flow solvers, it is possible to compute a $(1+\eps)$-approximation of the {\em cost-value} in $\widetilde O(n^{3/2})$ or in $\widetilde O(n^{4/3+o(1)})$ time, also supporting point multiplicity (see \cite{gabow1989faster} and \cite{axiotis2020circulation}).
See also e.g.,  \cite{esa/CabelloGKR05,comgeo/CabelloGKR08} and  \cite{gudmundsson2007small}.
\cite{soda/Indyk07} showed that importance sampling can be used for obtaining, with high probability, an $O(1)$-approximation of the cost-value in $O(n \operatorname{poly}(\log n))$ time, which however also does not allow to report a matching. \cite{altschuler2017near} presented an algorithm that also reports a matching, but with an additive error \(\epsilon\) in time \(\tilde{O}(n^2/\epsilon^3)\). 
For the parallel and streaming model, \cite{stoc/AndoniNOY14} presented an $n^{1+o_\eps(1)}$ time algorithm for $\mathbb{R}^2$ that allows to compute an $O(1)$-approximation for the cost-value of an optimal matching.
\cite{soda/SharathkumarA12} also presented a dynamic approximation algorithm that maintains a partial-matching. 

\section{Formal description of the implicit and explicit matching sub-routines}

\label{sec:app:implicit-explicit}

\paragraph{\textsc{Implicit Matching}}

Recall that the input of the procedure is a vertex \(v\) of \(p\)-tree \(T\) and the monochromatic set of points \(\mathcal{E}:=\{E_u:u\in \mathbb{C}_v\}\) forwarded to \(v\) from its children and their cardinalities \(\{|E_u|:u\in \mathbb{C}_v\}\). 

\textit{Choosing the points to match at v.} Formally, first divide \(\mathcal{E}\) into collections of red and blue monochromatic sets \(\mathcal{E}_1:=\{E_u:u\in \mathbb{C}_v, E_u \text{ is red} \}\) and \(\mathcal{E}_2:=\{E_u:u\in \mathbb{C}_v, E_u \text{ is blue} \}\), respectively. 

Then, determine the total number of red and blue points in \(\mathcal{E}_1\) and \(\mathcal{E}_2\), respectively, by summing up the provided set cardinalities in each collection. Suppose that there are $k_1$ red points in \(\mathcal{E}_1\), $k_2$ blue points in \(\mathcal{E}_2\), and that without loss of generality $k_1\leq k_2$. In this case, we aim to match all red points from sets in \(\mathcal{E}_1\). To pick $k_1$ points from blue sets in $\mathcal{E}_2$, we loop over $\mathbb{C}_v$ in a fixed order (e.g. z-order w.r.t. the underlying \(p\)-tree) 
and pick all elements $E_u\in \mathcal{E}_2$ until one child $u$ provides sufficiently many points to satisfy stopping (i.e. picking all \(|E_u|\) points would lead to a set of blue points of size at least $k$).

From this child, the selection implicitly chooses the remaining set \(E'_u\) of $k'\leq k_1$ points with the smallest $z$-order. Define set \(E_v:=E_u \setminus E'_u\) of points that will not be matched at \(v\), which can be represented by  an integer $k$ and a pointer to child $u$ a pointer to the \((k'+1)\)-th element in set \(E_u\). 
Finally, denote the set of chosen red and blue points by \(Y_v(A)\) and \(Y_v(B)\), respectively, and let \(Y_v=Y_v(A)\cup Y_v(B)\).
Note that (implicitly) choosing these subsets at a node $v$ takes $O(p^2)$ time, though reporting the points in $Y_v(A)$, or in $Y_v(B)$, takes $O(k\cdot d)$ time, where $d=O(\log_p U)$ is the depth of the $p$-tree.

\textit{Obtaining the aggregated sub-instance.} For any cell $u \in \mathbb{C}_{v}$, define $z_u$ to be the \emph{center} point of $u$. 
For $x \in Y_v$, let $u(x)$ be the subcell of $\mathbb{C}_v$ that $x$ is contained in.
We define a mapping $m(x)$ from \(x\) to $z_{u(x)}$, to be the function that maps a point $x$ to the center of its (bounding) subcell. 

For any point set $S$, let $m(S) := \{m(x) \mid x \in S \}$. 
We call $m(Y_v)$ a \emph{aggregated} sub-instance w.r.t. to node $v$ and the corresponding point set $Y_v$. 

Note that the function $m(Y_v)$ can be implicitly represented by the $p^2$ sub-grid cells of the cell of $v$ and their corresponding points in $Y_v$. 

\textit{Computing the implicit matching:}
We solve the following Euclidean transport instance on the points of $m(Y_v(A)), m(Y_v(B))$: For every point $a' \in m(Y_v(A))$, 
we set the supply of $a'$ to be the number of points in $Y_v(A)$ that map to $a'$, i.e., $s_{a'} := |\{a \in Y_v(A) : m(a) = a'\}|$.
Similarly, for every point $b' \in m(Y_v(B))$, we set the demand of $b'$ to be the negative number of points in $Y_v(B)$ that map to $b'$, i.e., $d_{b'} := -|\{b \in Y_v(B) : m(b) = b'\}|$. Let \(\gamma^*_v[Y_v]\) be an optimal assignment (or implicit matching) obtained by solving the Euclidean transportation problem on the sets $m(Y_v(A))$ and $m(Y_v(B))$ using the algorithm of Theorem~\ref{thm:transport}.

\paragraph{\textsc{Explicit Matching}} The inputs to the explicit matching procedure are a vertex of the restricted p-tree $v$, the point sets $Y_v$ (partitioned by their corresponding cells in $u \in \mathbb{C}_{v}$ and an optimal assignment $\gamma^*_v[Y_v]$ to the Euclidean transport instance on $(m(Y_v(A)), m(Y_v(B)),s,d)$ described above. Its output is a matching of points $Y_v$.

The procedure iterates through the edges $(a',b') \in m(Y_v(A))\times m(Y_v(B))$ with non-zero assignment in $\gamma^*_v[Y_v]$ in an arbitrary manner and adds $\gamma^*_v[Y_v](a',b')$ arbitrary pairs of points $a,b$, where  $a \in Y_v(A)$ and $b \in Y_v(B)$, and $m(a) = a'$  and $m(b) = b'$, as edges to the output matching from the cells with centers $a',b'$.

\section{Proof of Theorem~\ref{thm:static}}
\label{sec:app:static-proof}

\static*

\begin{proof}[Proof of Theorem \ref{thm:static}]
For this proof, we will assume that both \(n\) and \(p\) are powers of $2$.
While a similar approach extends to the general case, we present the special case here for the sake of clarity.
We show that \(p=n^{\eps /10}\) yields the runtime and approximation bounds.
    
    {\bf \em Run-time analysis: }
    By definition each leaf of \(p\)-tree \(T\) contains at most $p^2$ 
    many points.
    Now, since, we can compute the exact optimal matching between the points $Y_v$ for leaf node $v$ in $O(|Y_v|^3)$ time using the Hungarian algorithm, the total runtime of computing exact matchings of all leaves is bounded by $O(p^4 \cdot n)$. 

    Next, we discuss the running time of step~\ref{implicit_step} in \(\textsc{Implicit-Matching}(v, Y_v)\). 
    By Lemma \ref{lm:implicitMatching}, for each non-leaf node $v$, the \textsc{Implicit-Matching} in the execution of $\match(v)$ can be implemented in $O(p^2+p^5 \cdot \log^2 n)=O(p^5 \cdot \log^2 n)$ time.
    In addition, we need to solve \textsc{Implicit-Matching} for at most $n$ nodes in the $p$-tree \(T\).
    The total run-time of solving the \textsc{Implicit-Matching} at each node is bounded by $O(n\cdot p^5 \cdot  \log^2 n)$. 
    
    For any node $v$, the total time to report the points in excess set $E_v$ is bounded by $O(|E_v|)$.
    Thus, Lemma~\ref{lemma:explicit} yields the perfect matching on set \(X\) takes \(O(n)\) time.
    Hence, the runtime of \textup{\textsc{Static-Matching}$(A,B)$} is at most
    $O \left (n+p^4 \cdot  n + n\cdot p^5 \cdot \log^2 n + n \cdot \frac{\log U}{\log p} \right)$,
    since the number of levels in \(T\) is bounded by $\frac{\log U}{\log p}$. 
    Setting $p = n^{\epsilon/10}$, 
    and using that \(U\leq n^c\) for a fixed \(c>0\), we obtain the stated run-time bound.

    Note that while the upper bound we present of the approximation ratio holds in expectation, in the static setting we may repeat the algorithm $O(\log n)$ times and output the smallest cost solution to find an $O(1/\epsilon)$-approximate solution with high probability. In the dynamic setting, this simple trick guaranteeing a high-probability approximation ratio does not work.

    {\bf \em Correctness analysis:} 
    Let \(\mu^*\) be a matching on set \(X\) with optimal cost $c(\mu^*)$. Say that an edge is crossing at level $i$ if it crosses the grid of level $i$ of $T$. Observe that an edge may be crossing at multiple levels. Let $\cross(M,i)$ stand for the number of crossing edges of matching $M$ at level $i$.
    
    
    Note that if none of the edges of $\mu^*$ is cut by a grid at level $\ell$, this implies that all nodes at level $\ell$ have a balanced number of blue and red points. Also note that no edge of $\mu^*$ is crossing at level $0$, as we assume input points are contained in the cell $D \times D$ for some $D = n^c$ (both in the static and dynamic settings).
    

    
    Let \(\mu_a\) be the matching computed with our algorithm w.r.t. \(T\). 
    Our goal is to prove that 
    \begin{equation} \label{toprove}
        \mathbb{E}[c(\mu_a)]= O({\varepsilon}^{-1}\cdot c(\mu^*))\quad ,
    \end{equation} 

where the expectation is taken with respect to the random shift introduced at initialization. Let $T_\ell$ stand for the vertices of $T$ at level $\ell$. Let $\lambda_\ell = D/p^\ell$ stand for the length of the sides of the grid at level $\ell$. Slightly overloading notation let $\lambda_v$ stand for the length of side of the cell of $v$.
    
Matching $\mu_a$ has the following convenient property:
every point of $Y_v$ is matched with a point of $Y_v$ (inside the cell of $v$) under $\mu_a$ due to the construction of our implicit-explicit matching procedure.
Unfortunately, $\mu^*$ might not have this convenient property, making the comparison between them difficult. 
We now describe a procedure which translates $\mu^*$ into a matching $\mu'$ with this property. 
Later, we compare the costs of $\mu',\mu^*$ and $\mu_a$.

    \paragraph{Generating $\mu'$ from $\mu^*$}

    We describe an iterative method which generates $\mu'$ from $\mu^*$. 
    Initially, we set $\mu' = \mu^*$. 
    Assume that the tree $T$ has $k$ levels. Iterating $\ell$ from $k$ to $0$ for every $v \in T_\ell$, we keep repeating the following as long as possible:
    
    If there exists edges of $\mu'$, denoted $(r_1,b_1),(r_2,b_2)$, such that $r_1,b_2 \in Y_v$ for some $v \in T_\ell$, $r_1,b_2$ are of opposite color and $b_1,r_2 \notin Y_v$, then replace $(r_1,b_1),(r_2,b_2)$ with $(r_1,b_2),(r_2,b_1)$ in $\mu'$.
        

    Let $\mu_\ell'$ for $\ell \in \{k, \dots, 0\}$ stand for the state of $\mu'$ after not being able to repeat this process at level $\ell+1$. Specifically, $\mu^* = \mu_{k}'$.

    \begin{claim}
    \label{cl:static-proof:termination}

    The above process terminates in $n$ iterations and for any $\ell$ it must hold that for all $v \in T_{> \ell}$ the points of $Y_v$ are matched exclusively to each other by $\mu_\ell$.

    \end{claim}
    \begin{proof}

    Observe that before the execution of any of the two steps none of the two edges run between points of $Y_w$ for any $w \in T$. However, after the execution we have $(r_1,b_2) \in \mu'$ and $r_1,b_2 \in Y_v$ for some $v \in T$. Hence, the process must terminate in $n$ steps.

    Also, note that any $Y_v$ contains the same number of red and blue points. If, without loss of generality, any red point of some $Y_v$ is matched outside of $Y_v$ by $\mu'$, then there must also exist a blue point of $Y_v$ matched outside of $Y_v$ and the process can take a step. Hence, once the process can't make any further steps for some $Y_v$ where $r_1,b_2 \in Y_v$, then $\mu'$ is matching the points of $Y_v$ to each other.
    \end{proof}

    \begin{claim}
    \label{cl:static-proof:crossing-edges}

    $\cross(\mu_\ell',\ell') \leq \cross(\mu^*,\ell')$ for all $\ell' < \ell$.    
    \end{claim}

    \begin{proof}

    Assume that the process is progressing through vertices of level $\ell$. We will argue that any step it may take will not increase the number of crossing edges at lower levels, implying the claim. If the process takes a step for some $v \in T_\ell$ then it adds edges $(r_1,b_2),(r_2,b_1)$ to $\mu'$. $(r_1,b_2)$ is in the same cell at level $\ell$ hence it is not crossing at lover levels. $(r_2,b_1)$ might be crossing at some level $\ell' < \ell$. However, in this case as $(r_1,b_2)$ are in the same cell of level $\ell'$ it must hold that in this case either $(r_1,b_1)$ or $(r_2,b_2)$ was crossing at level $\ell'$ before their removal from $\mu'$.

    \end{proof}

    We will first relate the costs of matchings $\mu^*$ and $\mu'$.

    \begin{claim}
    \label{cl:static-proof:m1m*}

    $c(\mu') \leq c(\mu^*) + \sum_{\ell = 0}^k 8\cdot \lambda_\ell \cdot \cross(\mu^*,\ell)$.

    \end{claim}

    \begin{proof}

Consider the state of $\mu'$ after having finished with vertices of level $\ell+1$, that is $\mu_{\ell}'$. Call a point \textit{bad} at this point if it belongs to some $Y_v : v \in T_\ell$ and is matched to some point outside of $Y_v$ by $\mu_{\ell}'$. In this case there are two options: it is either matched outside of its cell at level $\ell$, or it is matched to a point forwarded by node $v$ to its ancestor but still inside $\mathcal{C}_v$. Note that it can't be matched to any point of $Y_w$ for a node on higher level then $v$ due to Claim~\ref{cl:static-proof:termination}.

The set of points forwarded by $v$ to its parent is monochromatic. Hence, for any bad point of the later type we will be able to associate a bad point of the earlier type. This implies, that at least half of the bad points, denote their number by $\mathcal{B}_\ell$, are incident on edges crossing at level $\ell$ in $\mu_{\ell}’$, that is $\mu_{\ell}’$ has at least $\mathcal{B}_\ell/2$ crossing edges at level $\ell$. Observe that, due to Claim~\ref{cl:static-proof:crossing-edges}, this implies that $\mu^*$ also has at least $\mathcal{B}_\ell/2$ crossing edges at level $\ell$.
Once the process gets to level $\ell$ in every step, it will reduce the number of bad edges by $1$. Furthermore, we claim that in each step it may increase the cost of $\mu’$ by at most $4 \cdot \lambda_\ell$. By every step some $(r_1,b_2),(r_2,b_1)$ are added to $\mu’$ and some $(r_1,b_1),(r_2,b_2)$  are removed. We have that $c(r_1,b_2) \leq 2 \cdot \lambda_\ell$ as $r_1$ and $b_2$ are contained in the same cell at level $\ell$. By triangle inequality, we have $c(r_2,b_1) \leq c(b_1,r_1)+c(r_1,b_2)+c(b_2,r_2)$ and therefore $c(r_2,b_1)+c(r_1,b_2) - c(b_1,r_1) - c(b_2,r_2) \leq 2\cdot c(r_1,b_2) $. Hence the total increase is indeed at most $4 \cdot \lambda_\ell$.
That is, the process will take at most $B_\ell$ steps, in each increasing $\mu’$ by at most $4 \cdot \lambda_\ell$, where $\mathcal{B}_\ell \leq \cross(\mu^*,\ell)/2$. Summing over all levels proves the claim.

\end{proof}

We now relate the costs of matchings $\mu'$ and $\mu_a$. Let $M[Y]$ stand for the edges of matching $M$ restricted to points $Y$. By construction of the algorithm and due to Claim~\ref{cl:static-proof:termination}, we know that both of $\mu’$ and $\mu_a$ match the points of $Y_v$ to each other for any $v \in T$. First, observe that if $v$ is a leaf then our algorithm calculated an optimal minimum cost matching of $Y_v$, that is $c(\mu'[Y_v]) \leq c(\mu_a[Y_v])$ for all leafs $v \in V$.

    \begin{claim}
\label{cl:static-proof:mamdash}
    $c(\mu_a[Y_v]) \leq c(\mu'[Y_v]) + 2\cdot\lambda_{v}/p \cdot |Y_v|$ for all non-leaf $v \in T$.
    \end{claim}
    \begin{proof}

    Recall that the algorithm computes $\mu_a[Y_v]$ by aggregating all points of $Y_v$ to the respective center of their sub-cell in $\mathbb{C}_v$ and computing an optimal solution of the resulting Euclidean geometric transport instance, then moving the points back to their original location. That is, assuming the points were already in the center of their respective cells, $\mu_a[Y_v]$ is a minimum cost matching of $Y_v$. Moving a point to the center of it's respective sub-cell increases the cost of an optimal solution by at most $\lambda_{v}/p$ (the length of the sides of the sub-cells of $v$), similarly the cost of moving them back to their original position is also upper bounded by $\lambda_{v}/p$.

    \end{proof}

    \begin{claim}

\label{cl:static-proof:mdashmstar}

    $\sum_{v \in T_\ell | v \textit{ is non-leaf}} |Y_v| \leq 2 \cdot \cross(\mu^*,\ell+1)$ for $\ell \in \{0 \dots k-1\}$.

    \end{claim}

    \begin{proof}

    Consider a particular vertex $v \in T_\ell$. The points matched at $v$ (the points of $Y_v$) are all inherited by $v$ by its children nodes due to an imbalance of red and blue points in the respective cells of $v$-s children nodes. This implies that $\mu^*$ matched at least $|Y_v|$ points in $\mathbb{C}_v$ with points in a separate sub-cell of layer $\ell+1$. Hence, $|Y_v|$ points of $\mathbb{C}_v$ are incident on crossing edges at level $\ell+1$ in $\mu^*$. Summing over all vertices of $\{Y_v | v \in T_\ell\}$ proves the claim.

    \end{proof}

    We now have to tie the number of crossing edges to the cost of $\mu^*$.

    \begin{claim}

    \label{cl:static-proof:crosstomstar}

    $\mathbb{E}[\cross(\mu^*,\ell)]/\lambda_\ell = O( c(\mu^*))$ for all $\ell \in \{0 \dots k\}$ where the expectation is taken with respect to the random shift introduced at initialization.

    \end{claim}

    \begin{proof}

    Consider an edge $e \in \mu^*$ of length $L$. With probability $\Omega(L/\lambda_\ell)$ it crosses the grid at level $\ell$ due to the random shift. Hence, it contributes $L$ to both sides of the inequality.
        
    \end{proof}

     We are now ready to upper bound the approximation ratio of the algorithm in expectation with respect to the random shift at initialization.

\begin{align}
\mathbb{E}[c(\mu_a)]  \quad= & \quad \sum_{v \in T | v \textit{ is leaf}} \mathbb{E}[c(\mu_a[Y_v])] +  \sum_{v \in T | v \textit{ is non-leaf}} \mathbb{E}[c(\mu_a[Y_v])] \nonumber \\
 \quad \leq & \quad \mathbb{E}[c(\mu_1)] + \sum_{v \in T | v \textit{ is non-leaf}} 2 \cdot \mathbb{E}[|Y_v|] \cdot \lambda_v/p \label{eq:static-proof:1} \\
 \quad \leq & \quad c(\mu^*) + \sum_{\ell = 0}^k 8 \cdot \lambda_\ell \cdot \mathbb{E}[\cross(\mu^*,\ell)] + \sum_{v \in T | v \textit{ is non-leaf}} 2 \cdot \mathbb{E}[|Y_v|] \cdot \lambda_v/p \label{eq:static-proof:2} \\
 \quad \leq & \quad c(\mu^*) + \sum_{\ell = 0}^k 8 \cdot \lambda_\ell \cdot \mathbb{E}[\cross(\mu^*,\ell)] + \sum_{\ell = 0}^{k-1} 4 \cdot \lambda_\ell/p \cdot \mathbb{E}[\cross(\mu^*,\ell+1)] \label{eq:static-proof:3} \\
 \quad \leq & \quad O(c(\mu^*) \cdot k) \label{eq:static-proof:4} \\
 \quad \leq & \quad O(c(\mu^*)/\epsilon) \nonumber
\end{align}

The first inequality follows from the fact that $\mu_a$ matches vertices of $Y_v$ exclusively with each other. Equation~\ref{eq:static-proof:1} holds due to Claim~\ref{cl:static-proof:mamdash}. Equation~\ref{eq:static-proof:2} follows from Claim~\ref{cl:static-proof:m1m*}, while equation~\ref{eq:static-proof:3} follows from Claim~\ref{cl:static-proof:mdashmstar}.

Equation~\ref{eq:static-proof:4} follows from Claim~\ref{cl:static-proof:crosstomstar}. Finally, the last inequality holds as the number of layers in the $p$-tree $T$ is upper bounded by $O(\log U/ \log P) = O(1/\epsilon)$.

\end{proof}

\section{Basic Algorithm}
\label{app:basic}

In this section we first present the pseudo-code of our basic algorithm which can answer queries about the solution efficiently (see Theorem~\ref{thm:basic}). Afterwards we argue about it's running time and approximation ratio. 

\begin{algorithm}
\caption{$\textsc{Insert-Update}(v, a)$}\label{alg:insert-update}
\begin{algorithmic}[1]

\If{\(v\) is non-marginal internal node of \(T\)} 
    \State{\(u \leftarrow\) child $u \in \mathbb{C}_v$ such that \(a\in X_u\)} 
    \State{Update \(E_u\) and \(c(\mu_u)\) via \(\textsc{Insert-Update}(u, a)\)}
    \State{Update \(\gamma_v^*[Y_v]\) and \(E_v\) via \(\textsc{Implicit-Matching}(v, \mathcal{E}, \{|E_u|: u \in \mathbb{C}_v\})\)}
    \State{\(c(\mu_v) \leftarrow \sum_{u\in \mathbb{C}_v} c(\mu_u)+c(\gamma_v^*[Y_v])\)}
\EndIf

\If{\(v\) is non-marginal leaf in \(T\)}
    \State{\(Y_v'(A) \leftarrow Y_v'(A)\cup \{a\}\)}
    
    \If{$\eta\geq  0$}  
    
        \State{\(E_v \leftarrow E_v\cup \{a\}\)}
        \State{$\eta \leftarrow \eta + 1$}
        
        \State{\(Y_v(A) \leftarrow Y_v(A) \setminus E_v\)}
    \Else
        \State{\(b \leftarrow\) any point from \(E_v\)}
        \State{\(E_v \leftarrow E_v\setminus \{b\}\)}
        \State{\(\eta \leftarrow \eta - 1\)}         

        \State{\(Y_v(B) \leftarrow Y_v(B) \setminus E_v\)}
    \EndIf
    \State{\(\mu_v \leftarrow \textsc{Hungarian-Algorithm}(Y_v(A), Y_v(B))\)}
\EndIf

\If{\(v\) is a new subroot of \(T\)}

    \If{$\eta\geq  0$}
    
        \State{\(E_v \leftarrow E_v\cup \{a\}\)}
        \State{$\eta \leftarrow \eta + 1$
        }
        
    \Else
        \State{\(b \leftarrow\) any point from \(E_v\)}
        \State{\(E_v \leftarrow E_v\setminus \{b\}\)}
        \State{\(\eta \leftarrow \eta - 1\)}
        
    \EndIf
    \State{Update \(\gamma_v^*[Y_v]\) and \(c(\mu_v)\) via \(\textsc{Match}(v)\)}
    \State{\(c(\mu_v) \leftarrow \sum_{u\in \mathbb{C}_v} c(\mu_u)+c(\gamma_v^*[Y_v])\)}
\EndIf

\end{algorithmic}
\end{algorithm}

\begin{algorithm}
\begin{algorithmic}[1]
\caption{$\textsc{Insert}(a)$}\label{alg:insert}
    \State{Update restricted \(p\)-tree \(T\)}
    \State{\(\textsc{Insert-Update}(r, a)\)}

\end{algorithmic}
\end{algorithm}

\begin{algorithm}

\caption{$\textsc{Delete-Update}(v, a)$}\label{alg:delete-update}
\begin{algorithmic}[1]
\If{\(v\) is an internal node of \(T\)}
    \State{\(u \leftarrow\) child $u \in T_v$ such that \(a\in X_u\)}
    \State{Update \(E_u\) and \(c(\mu_u)\) via \(\textsc{Insert-Update}(u, a)\)}
    \State{Update \(\gamma_v^*[Y_v]\) and \(E_v\) via \(\textsc{Implicit-Matching}(v, \mathcal{E}, \{|E_u|: u \in \mathbb{C}_v\})\)}
    \State{\(c(\mu_v) \leftarrow \sum_{u\in \mathbb{C}_v} c(\mu_u)+c(\gamma_v^*[Y_v])\)}
\EndIf

\If{\(v\) is a leaf in \(T\)}
    \State{\(Y_v'(A) \leftarrow Y_v'(A)\setminus \{a\}\)}
    
    \If{\(a\in E_v\)}
        \State{\(E_v \leftarrow E_v\setminus \{a\}\)}

    \Else
        \If{\(\eta >  0\)}

            \State{\(b \leftarrow\) any point from \(E_v\)}
            \State{\(E_v \leftarrow E_v\setminus \{b\}\)}
            \State{\(\eta \leftarrow \eta - 1\)} 
            
            \State{\(Y_v(B) \leftarrow Y_v(B) \setminus E_v\)}
        \Else
            
            \State{\(E_v \leftarrow E_v\cup \{a\}\)}
            \State{\(\eta \leftarrow \eta + 1\)}
            
            \State{\(Y_v(A) \leftarrow Y_v(A) \setminus E_v\)}    
        \EndIf
        \State{\(\mu_v \leftarrow \textsc{Hungarian-Algorithm}(Y_v(A), Y_v(B))\)}
    \EndIf
\EndIf

\end{algorithmic}
\end{algorithm}

\begin{algorithm}
\begin{algorithmic}[1]
\caption{$\textsc{Delete}(a)$}\label{alg:delete}
    \State{Update restricted \(p\)-tree \(T\)}
    \State{\(\textsc{Delete-Update}(r, a)\)}

\end{algorithmic}
\end{algorithm}

\subsection{Running Time Analysis}

We will first argue about the update time of the algorithm. We will analyze the update time of an insertion as the process of handling of deletions is similar. An update is handled in a top-to-bottom manner: the algorithm first calls the update procedure at the root and then always recuses towards the child which contains the newly inserted point. Hence, throughout an update the algorithm will visit nodes of $T$ along a path form the root to some leaf. Apart from the visited leaf node the only non-constant time operation executed at any node is the implicit-matching procedure which runs a geometric Euclidean transport sub-routine on an aggregated sub-instance.

These aggregated sub-instances consist of points corresponding to the centers of the sub-cells of the respective node, that is $p^2$ points. The weights assigned to each center are at most $n$. Hence, by Theorem~\ref{thm:transport} they run in time $O(p^5 \cdot \log (p \log n)) = O(n^{O(\epsilon)})$. That is, updating the non-leaf nodes along the path takes $O(n^{O(\epsilon)}/\epsilon)$ worst case update time, as $T$ will always have at most $O(1/\epsilon)$-levels.

The leaf node might end up having more then $p^2$ input points in its cell after the insertion (otherwise we simply call the Hungarian algorithm on it). This might result in the leaf splitting into a new sub-tree. This sub-tree can similarly only have $O(1/\epsilon)$-levels and in total at most $O(p^2/\epsilon)$ nodes. On each of its non-leaf nodes similarly the only non-constant time operation will be the implicit matching procedure adding $O(n^{O(\epsilon)}/\epsilon)$ worst-case update time. 

There might be $\sim p^2$ leaves of the new sub-tree, but the number of points inside the cells of the new leaves is at most $O(p^2)$. In each such leaf the only non-constant time calculation the algorithm completes is a call to the Hungarian algorithm to obtain an optimal solution inside the leaf, hence this contributes at most $O(p^6 = n^{O(\epsilon)})$ to the update time. 

Additionally, we can maintain the p-tree $T$ itself in $O(p^2/\epsilon) = O(n^{O(\epsilon)}/\epsilon)$ worst-case update time. Hence, the total worst-case update time of the algorithm is $O(n^{O(\epsilon)}/\epsilon)$.

\subsection{Algorithm Correctness}

\label{app:sec:algo-croectness}

The analysis of the approximation ratio of the static algorithm presented in Section~\ref{sec:app:static-proof} relies on the following properties of the output: 

\begin{enumerate}
    \item[(1)] For any node $v \in T$ points of $Y_v$ (which all lie in the cell of $v$) are matched exclusively to each other.
    \item[(2)] For any leaf node $v \in T$ the points matched in the cell of $v$, $Y_v$, are matched via an optimal Euclidean bi-chromatic matching algorithm (Hungarian algorithm).
    \item[(3)] For any internal node $v \in T$ the points matched in the cell of $v$, $Y_v$ are matched via the implicit-explicit matching procedures relying on an underlying optimal Euclidean geometric transport sub-routine (Theorem~\ref{thm:transport}).
    \item[(4)] For any node $v \in V$ the algorithm matches all but a monochromatic set of points out of the sets of points forwarded to $v$ by its children inside the cell of $v$ (or in the case of leaves among all the points in the corresponding cell $\mathcal{C}_v$).
\end{enumerate}

We will argue that once an update is processed by the dynamic algorithm, all of these properties of the output are restored. The only vertices of $T$ where any of the mentioned properties might be violated are the ones whose cell contains the updated point or ones which have been created due to the update to the $p$-tree. 

Leaves, which have contained the updated point or are created due to the update, all select an arbitrary maximal color balanced set of points in their respective cell. The algorithm runs the Hungarian algorithm on these sets restoring item (2). Afterwards, leaves pass on a monochromatic set of points to their parent internal nodes.

Every internal node runs the implicit-explicit matching procedure on the points passed to it by its children in a bottom-up order. This implies they will satisfy item (3) once the update has been processed. The implicit-explicit matching procedure matches all but a monochromatic set of points in a cell, which set is then forwarded to the parent node. This implies that by the time the root has finished its internal calculations, both items (4) and (1) are satisfied.

Note that we may guarantee that the approximation ratio holds with high probability if we simply maintain $O(\log n)$ solutions in parallel and always query the one with smallest cost. However, in this case in order to maintain the approximation ratio with high probability over super-polynomially long update sequences the data-structure has to be periodically rebuilt reducing the update time guarantee to be amortized.

\section{Advanced Algorithm}

\label{sec:app:advanced}

\label{app:update}


 In this section, we describe an improvement of the algorithm described in Section \ref{dynamic}.
The main difference is that in the advanced algorithm, we maintain an explicit matching at each node. As shown previously, we may recompute the explicit matching at every node efficiently after an update through a black-box Euclidean geometric transport algorithm. However, without opening the black-box, we can't guarantee that the implicit matching computed will not differ significantly from its previous version. This means that in order to maintain a matching of the actual input points we would have to re-run the explicit matching procedure, but that takes time proportional to the number of points matched at the specific node which could be $\Omega(n)$. 

To overcome this difficulty, we show that we can update the implicit and explicit matchings stored at each node $v \in T$ in time $O(p^3)$ when two points are added to the respective point set $Y_v$. We will call this procedure \(\textsc{Augment-Matching}\). We then show that the size of the excess set of each node of $T$ changes by exactly $1$ after each update. We finally present an algorithm which maintains the explicit matchings stored at each node efficiently.


\subsection{Updating Implicit/Explicit Matchings Under Insertions} 

Let $\mu$ be a matching of some red and blue point sets $(A,B)$ and $x,y$ be some arbitrary points of opposite color not in $A,B$. Define an augmenting path with respect to $\mu$ to be a path between $x,y$ consisting of edges between $x,y,A,B$, alternating between edges not in $\mu^*$ and in $\mu^*$. Define the cost of such a path to be the sum of the costs of its edges not in $\mu$ minus the cost of its edges in $\mu$. 

For such an augmenting path $\Pi$, define the matching $\mu'$ consisting of all of the edges of $\mu$ not in $\Pi$ and the edges of $\Pi$ not in $\mu$ to the matching $\mu$ augmented by $\Pi$. Observe that $c(\mu') = c(\mu) + c(\Pi)$ and $c(\mu')$ is a perfect matching of $A \cup \{x\}, B \cup \{y\}$.

\begin{claim}

\label{cl:advanced-algo:augmentation}

Let $A,B$ be a set of input points for the bi-chromatic Euclidean matching problem and $\mu^*$ be an optimal matching of points $A,B$. Let $x,y$ be some arbitrary points in the plane of opposite colors and let $\Pi^*$ the smallest cost augmenting path with respect to $\mu^*$ starting and finishing at $x$ and $y$, respectively. Then the matching obtained by augmenting $\mu^*$ by $\Pi^*$ is a minimum cost perfect matching of $A \cup \{x\}, B \cup \{y\}$.

\end{claim}

\begin{proof}

Define $(\mu^*)'$ to be the matching obtained by augmenting $\mu^*$ with $\Pi^*$. Assume that it's not of minimum cost and let $\mu_E^*$ stand for a min cost matching of the extended point set $A \cup \{x\}, B \cup \{y\}$. Look at the union of $\mu^*$ and $\mu_E^*$. It consists of disjoint edges, cycles and a single augmenting path $\Pi^*$ with respect to $\mu^*$ between $x$ and $y$. Observe, that the cost of matching all points not part of $\Pi'$ is the same under both matchings due to their optimality in their respective point sets, hence $c(\mu_E^*) = c(\mu^*) + c(\Pi')$ . However, as $c((\mu^*)') = c(\mu^*) + c(\Pi^*)$ this would imply that $c(\Pi^*) > c(\Pi')$ which is a contradiction.

\end{proof}

Now extend the problem bi-chromatic Euclidean matching to point sets where certain points might have the same coordinates. It can be thought as extending the problem to inputs where input points have multiplicities.

\begin{claim}

\label{cl:advanced-algo:shortpath}

Let $A,B$ a set of red and blue points where some pairs of points might share locations and $\mu^*$ be a min cost bi-chromatic Euclidean matching between them. Assume that points $A,B$ are located on $k$ disjoint points of the plane, and points of $A\cup B$ at the same location are monochromatic. For arbitrary red and blue points $x,y$, there exists a min cost augmenting path between $x$ and $y$ with respect to $\mu^*$ which visits each of the $k$ disjoint locations at most once.

\end{claim}

\begin{proof}

For sake of contradiction assume that there is no such min cost augment path and let $\pi^*$ be a minimum length (in terms of number of edges) min cost augmenting path between $x$ and $y$ with respect to $\mu^*$. We will show that we can `shortcut' $\pi^*$.

Assume that $\Pi^*$ visits one of the $k$ disjoint locations $a$ twice. Then it must be the case that the $\Pi^*$ contains a sub-path $C$ which is a cycle starting and ending with $a$ consisting of $\mu^*$ and non-$\mu^*$ edges. There are three possible cases.
\begin{itemize}
\item If the total cost of non-$\mu^*$ edges of $C$ is the same as that of the $\mu^*$ edges, then we may remove this cycle meaning that $\Pi^*$ is not of minimum length.

\item If the total cost of non-$\mu^*$ edges of $C$ is larger then that of the $\mu^*$ edges, then if this cycle is removed from the path, we would get a smaller cost augmenting path between $x$ and $y$.

\item If the total cost of non $\mu^*$ edges of $C$ is smaller then that of the $\mu^*$ edges, then the matching $(\mu^* \setminus C) \cup (C \setminus \mu^*)$ (the matching received by augmenting $\mu^*$ by $C$) is of smaller cost then $\mu^*$ and is a perfect matching of $A,B$.
\end{itemize}

\end{proof}

\begin{theorem}
\label{thm:negative-shortestpath}

\cite{bellman1958routing} Belman-Ford: There exists an algorithm which given a graph $G = (V,E)$ and distance function $c : E \rightarrow \mathbb{R}$ where $G$ doesn't contain a negative length cycle with respect to distance measure $c$, returns a shortest path between two vertices of $V$ in time $O(|V|\cdot|E|)$

\end{theorem}

\paragraph{Augment-Matching Procedure}

We are now ready to describe the procedure which augments the implicit matchings under vertex updates, while making sure only a small number edges are affected by the update.

Recall that an implicit matching $\gamma^*[Y_v]$ of $v \in T$ corresponds to a matching of the aggregated versions of points $Y_v$. Assume that points $(x,y)$ are inserted into $Y_v$. Define the following simple graph $G = (V,E)$ and corresponding edge distance function $c$: define $V$ to consist of $p^2$ nodes, $V_\mathcal{C}$ representing the center pints of the sub-cells of $v$, and two additional nodes $x',y'$. 

Without loss of generality assume $x$ and $y$ are red and blue respectively. Add an edge between $x'$ and all nodes of $v_\mathcal{C} \in V_\mathcal{C}$ to $E$ where $v$ corresponds to a sub-cell of $v$ containing only blue points in $Y_v$ (recall that the points of $Y_v$ in every sub-cell of $v$ are monochromatic). Set $c((x',v_\mathcal{C}))$ to correspond to the distance between the centers of sub-cells of $x'$ and $v_\mathcal{C}$.

Similarly, add an edge between $y'$ and all nodes of $v_\mathcal{C} \in V_\mathcal{C}$ to $E$ where $v$ corresponds to a sub-cell of $v$ containing only red points in $Y_v$, and set $c((x',v_\mathcal{C}))$ to correspond to the distance between the centers of sub-cells of $x'$ and $v_\mathcal{C}$.

In addition add an edge to $E$ between any two sub-cell center nodes $v_{\mathcal{C}},w_{\mathcal{C}}$ if their sub-cells contain points of the opposite color. Set $c((v_{\mathcal{C}},w_{\mathcal{C}}))$ to be the distance between the respective sub-cell centers if there isn't an edge between the centers in $\gamma^*[Y_v]$, and otherwise set it to be the negative of this distance.

By construction, a path between $x',y'$ in $G$ corresponds to an augmenting path of $\gamma^*[Y_v]$ between $x,y$. By Claim~\ref{cl:advanced-algo:shortpath}, the shortest $x',y'$ in $G$ corresponds to a minimum cost augmenting path between $x,y$ with respect to $\gamma^*[Y_v]$. Observe that a negative length cycle of $G$ could be used to reduce the cost of $\gamma^*[Y_v]$ which was assumed to be optimal, hence no such cycle should exist.

Finally, due to Claim~\ref{cl:advanced-algo:augmentation} we may augment $\gamma^*[Y_v]$ to be a min cost perfect matching of the aggregated versions of $Y_v \cup \{x,y\}$ with such a shortest $x',y'$ path.

Constructing graph $G$ may take $O(|V| + |E|) = O(p^2)$ time. We may find the shortest path in $O(p^3)$ time by Theorem~\ref{thm:negative-shortestpath}. Any shortest path in $G$ can be of length $O(p^2)$ as $G$ doesn't contain negative length cycles. Hence, we can update the implicit matching of $Y_v$ in $O(p^3)$ time. Once the the updated implicit matching is computed we can transform it to an explicit matching of $Y_v \cup \{x,y\}$ in time $O(p^2)$ since the difference between the old and updated state of $\gamma^*[Y_v]$ (and hence the corresponding explicit matching) is only $O(p^2)$ edges. We state this formally by the following lemma:

\begin{lemma} \label{lemma:augment1}
    Let \(T\) be a \(p\)-tree and \(v\) a node in \(T\), \(\gamma^*[Y_v]\) the implicit matching corresponding to node \(v\), and  \(x, y \notin Y_v\) two points of opposite colors. \(\textsc{Augment-Matching}(\gamma^*[Y_v], x, y)\) returns an optimal implicit matching of the aggregated versions of \(Y_v\cup \{x, y\}\), in \(O(p^3)\) time.
\end{lemma}

Note that while Lemma~\ref{lemma:augment1} can be easily be extended to point pair deletions, there is no need for this. Assume we want to update an implicit matching $\gamma^*[Y_v]$ when points $(x,y)$ are deleted from $Y_v$. Let the neighbors of $x,y$ under $\gamma^*[Y_v]$ be $x',y'$. We may remove edges $(x, x')$ and $(y, y')$ from $\gamma^*[Y_v]$ and re-insert point pair $(x',y')$ using \textsc{Augment-Matching} to simulate this deletion.





 \subsection{Dynamic Algorithm}
 
 We now describe \textsc{Initialize}, the update procedures \(\textsc{Insert}\) and \(\textsc{Delete}\), as well as \(\textsc{Query-Recourse}\). The procedure \(\textsc{Query-Cost}\) is the same as in the basic algorithm. We will only describe how to do point insertions, as the procedure for point deletions is almost identical. Similarly as before, given $p$-tree \(T\), our goal is to maintain sets $E_v, Y'_v(A), Y'_v(B)$, difference \(\eta_v\), matching $\gamma^*[Y_v]$, but also matchings $\mu'[Y_v]$ and $\mu_v$, together with its corresponding cost \(c(\mu_v)\) for all the nodes $v$ in \(T\) which is the key difference from the basic algorithm presented in the main body of the paper. We first begin by describing the initialization procedure. 



\paragraph{Initialization} The procedure \textsc{Initialize} is almost the same as in the basic version of the algorithm. However, in the advanced algorithm, at each node \(v\) of \(T\), we also convert the implicit matching to its corresponding explicit matching. This further allows us to maintain the complete matching of a subtree corresponding rooted at each of its nodes, defined as \(\mu_v: =\mu'[Y_v] \cup \bigcup_{u\in \mathbb{C}_v} \mu_u\), here $\mathbb C_v$ denotes the children of node $v$ in $p$-tree $T$.

\paragraph{Handling Insertions} 

Without loss of generality, we assume that we are inserting a red point \(a\) to \(A\). As in the basic algorithm, \(\textsc{Insert}\) proceeds in two steps: updating the restricted \(p\)-tree and updating the matchings at each affected node.
We define the recursive procedure \(\textsc{Insert-Update}(v, a)\) for three types of nodes in \(T\): new subroots, non-marginal leaf nodes, and non-marginal internal nodes.

If a leaf \(v\) is the new subroot of \(T\), the \(\textsc{Insert-Update}(v, a)\) procedure is the same as in the dynamic algorithm presented in Algorithm \ref{alg:insert-update}.

For non-marginal leaf nodes \(v\), the procedure \(\textsc{Insert-Update}(v, a)\) checks if there exists a pair of points of opposite color at \(v\), and if so, augments the matching at \(v\).

For non-marginal internal nodes \(v\), we have two cases: the excess set \(E_u\) of its affected child \(u\) either increases by one red point \(a'\), or decreases by one blue point \(b'\). In the first case, we proceed similarly as in the leaves. Namely, if there exists a pair of points of opposite color at \(v\), and if so, augments the matching at \(v\). In the second case, we first have to check if the blue point \(b'\) was a part of a matching at \(v\). If so, we first have to break that matching. After this, we proceed similarly to the first case.

Finally, we prove the following theorem:

\main*

\begin{proof}[Proof of Theorem~\ref{main}]
    As the proof of this theorem is very similar to the proof of Theorem \ref{thm:basic}, we only note the key differences between these two algorithms.\\
    \emph{Running time:} Compared to the basic algorithm, the running time of \textsc{Initialize} only increases by the time required to convert implicit matchings at nodes of \(T\) to their corresponding matchings of the input points, and therefore also store the complete matchings of the subtrees. By Lemma \ref{lemma:explicit}, this takes an additional \(O(n)\) time, making the total run time of the procedure still \(O(n^{1+\varepsilon})\). \\
    Now we discuss updates. As mentioned before, the difference between the basic and the advanced update algorithm is that now, instead of recomputing the entire implicit matching at each affected node (using \textsc{Augment-Matching}), we use \(\textsc{Augment-Matching}\) procedure to augment both the implicit and the aggregated matching, if needed. The computation of \textsc{Augment-Matching} takes \(O(p^3)\) time by Lemma \ref{lemma:augment1}, so we again obtain that \(O(n^\varepsilon)/\epsilon\) is running time required for the update procedure of this algorithm.\\
    Regarding queries, in the advanced algorithm, we can easily report the perfect matching by simply returning the complete matching of a subtree at root \(r\) of \(T\), that is \(\mu_r\). Further, we obtain a more efficient algorithm for reporting the recourse of the matching. Namely, note that the aggregated matching at each affected node changes only along the augmenting path found in \textsc{Augment-Matching}. Since the length of this path is at most \(p^2\), the total recourse is at most \(O(p^2\cdot \frac{\log U}{\log p})\), which is \(O(n^\varepsilon/\epsilon)\) for \(p=n^{\frac{\varepsilon}{10}}\).\\ 
    \emph{Correctness:} In terms of correctness of the approximation, the advanced and basic versions of the algorithms behave exactly the same. That is, the advanced algorithm also satisfies the following. For any $Y_v$ for $v \in T$ points of $Y_v$ are matched exclusively to each other. Further, for leaf $v \in T$ the matching of $Y_v$ is computed via an optimal algorithm, for any non-leaf $v \in T$ the matching of $Y_v$ is computed by the implicit-explicit matching procedures. Finally, for any cell of the grid, the number of points not matched within that cell exactly corresponds to the the color dis-balance of the points present in that cell (see the similar argument concerning the basic algorithm see Section~\ref{app:sec:algo-croectness}). As argued in the analysis presented in Section~\ref{sec:app:static-proof}, the algorithm correctness only relies on these properties of the output matching.
\end{proof}

\begin{algorithm}[H]
\caption{$\textsc{Insert-Update-1}(v, a)$}\label{alg:insert-update1}
\begin{algorithmic}[1]

\If{\(v\) is non-marginal internal node of \(T\)} 
    \State{\(u \leftarrow\) child $u \in \mathbb{C}_v$ such that \(a\in X_u\)} 
    \State{\(E_u^b \leftarrow E_u\)}
    \State{Update \(E_u\) via \(\textsc{Insert-Update}(u, a)\)}
    \If{\(E_u = E_u^b\cup \{a'\}\) for some \(a'\in A\)}
        \If{\(\eta_v\geq 0\)}
        
            \State{\(E_v \leftarrow E_v\cup \{a'\}\)}
        \Else
            \State{\(b' \leftarrow \) any point from \(E_v\)}
            \State{Update \(\gamma^*[Y_v]\) and \(\mu'[Y_v]\) via \textsc{Augment-Matching}\((\gamma^*[Y_v], a', b')\)}
        \EndIf
    \EndIf
    \If{\(E_u = E_u^b\setminus \{b'\}\) for some \(b'\in B\)}
        \If{\(\eta_v \geq 0\)}
            \State{\(a' \leftarrow \) point from \(A\) such that \((a', b')\in \mu'[Y_v]\)}
            \State{Delete \((a', b')\) from \(\mu'[Y_v]\)}
            \State{\(\gamma^*[Y_v](a', b') \leftarrow \gamma^*[Y_v](a', b') - 1\)}
            \State{\(E_v \leftarrow E_v\cup \{a'\}\)}
        \Else
            \If{\(b\in E_v\)}
                \State{\(E_v \leftarrow E_v \setminus \{b'\}\)}
            \Else
                \State{\(a' \leftarrow \) point from \(A\) such that \((a', b')\in \mu'[Y_v]\)}
            \State{Delete \((a', b')\) from \(\mu'[Y_v]\)}
            \State{\(\gamma^*[Y_v](a', b') \leftarrow \gamma^*[Y_v](a', b') - 1\)}
            \State{\(b'' \leftarrow \) any point from \(E_v\)}
            \State{Update \(\gamma^*[Y_v]\) and \(\mu'[Y_v]\) via \textsc{Augment-Matching}\((\gamma^*[Y_v], a', b'')\)}
            \State{\(E_v \leftarrow E_v\setminus \{b''\}\)}
            \EndIf
        \EndIf

    \EndIf

    \If{\(E_u = E_u^b\setminus \{b'\}\) for some \(b'\in B\)}
        \If{\(b'\) was matched in \(\gamma^*[Y_v]\) with some \(a'\in A\)}
            \State{\(b'' \leftarrow \text{ any point from } E_v\)}
            \State{Update \(\gamma^*[Y_v]\) and \(\mu'[Y_v]\) via \textsc{Augment-Matching}\((\gamma^*[Y_v], a', b'')\)}
        \EndIf
    \EndIf

\EndIf

\end{algorithmic}
\end{algorithm}

\begin{algorithm}[H]
\begin{algorithmic}[1]
 \setcounter{ALG@line}{38}

\If{\(v\) is non-marginal leaf in \(T\)}
    \State{\(Y_v'(A) \leftarrow Y_v'(A)\cup \{a\}\)}
    \State{\(\eta_v \leftarrow |Y_v'(A)|-|Y_v'(B))|\)}
    \If{\(\eta\geq  0\)}  
    
        \State{\(E_v \leftarrow E_v\cup \{a\}\)}

    \Else
        \State{\(b \leftarrow\) any point from \(E_v\)}
        \State{\(\mu_v \leftarrow \textsc{Augment-Matching}(\mu_v, a, b)\)}
        \State{\(E_v \leftarrow E_v\cup \{b\}\)}

    \EndIf
\EndIf

\If{\(v\) is a new subroot of \(T\)}

    \If{\(\eta\geq  0\)}
    
        \State{\(E_v \leftarrow E_v\cup \{a\}\)}
        \State{\(\eta \leftarrow \eta + 1\)}
  
    \Else
        \State{\(b \leftarrow\) any point from \(E_v\)}
        \State{\(E_v \leftarrow E_v\setminus \{b\}\)}
        \State{\(\eta \leftarrow \eta - 1\)}
        
    \EndIf
    \State{\text{Update } \(\gamma_v^*[Y_v]\) and \(c(\mu_v)\) via \(\textsc{Match}(v)\)}
        \State{\(c(\mu_v) \leftarrow \sum_{u\in \mathbb{C}_v} c(\mu_u)+c(\gamma_v^*[Y_v])\)}
    
\EndIf

\end{algorithmic}
\end{algorithm}

\begin{algorithm}[H]
\begin{algorithmic}[1]
\caption{$\textsc{Insert}(a)$}\label{alg:insert1}
    \State{Update restricted \(p\)-tree \(T\)}
    \State{\(\textsc{Insert-Update}(r, a)\)}

\end{algorithmic}
\end{algorithm}

\section{Hardness Bound for Partially Dynamic Bi-Chromatic Euclidean Matching}

\label{app:hardness}

In this section we present a simple construction which shows that even under point insertions no efficient dynamic algorithm may maintain an arbitrarily tight approximation to the bi-chromatic Euclidean matching problem. This is in contrast with the static setting, where multiple $(1+\epsilon)$-approximate near-optimal running time algorithms are known \cite{swat/AgarwalRSS22,jacm/RaghvendraA20}.

\begin{theorem}

\label{thm:hardness}

For any $\delta >0$, any  $(2 - \delta)$-approximate algorithm that maintains a solution to the Euclidean bi-chromatic matching problem under either insertions or deletions of pairs of points must have $\Omega(n)$ amortized update time.

\end{theorem}

\begin{proof}

For sake of contradiction, assume that a dynamic algorithm maintains a $(2-\delta)$-approximate solution for some constant $\delta > 0$ on some input $A,B$. We will consider the case of point pair insertions. Let the points of $A,B$ all lie on the $x$-axis. Assume that at some point the input consists of $2 \cdot n$ specific points for some $n >> 1/\delta$. We will show that processing the next $2 \cdot n$ points will take at least $\Omega(n^2)$ time proving the theorem.

Since the points lie on a line we will describe them with a single coordinate. Assume that the first $2\cdot n$ points of the input consists of $A = \{-n,-n+2,-n+4, \dots, n-2\}$, $B = \{-n+1, -n+3,\dots, n-1\}$ that is simply a series of points with alternating colors with unit distance between them.

The next $n$ updates will add a point a unit distance away to the left of the leftmost point and to the right to the rightmost point of the existing points, such that at all times the colors of the points is alternating. That is, the first update adds a point to $A$ at $n$ a point to $b$ at $-n-1$. Afterwards a point to $A$ at $-n-2$ and a point to $B$ at $n+1$ ect.

Let $\mu_a$ be the output of the algorithm at some point in time throughout these insertions, and let $\mu_a'$ be its output after processing the next insertions. Let the newly inserted points without loss of generality be $-x,x-1$.

In $\mu_a \cup \mu_a'$ points $-x,x-1$ are connected by a path of alternating edges of $\mu_a$ and $\mu_a'$, lets call this path $\Pi$. We will show that $\Pi$ consists of $\Omega(x) = \Omega(n)$. This implies that during the processing of input $(-x,x-1)$ the output of the algorithm undergoes at least $\Omega(n)$ recourse, lower bounding its update time.

For the sake of contradiction, assume that the length of $\Pi$ is some $o(x)$. The total cost of matchings $\mu_a$ and $\mu_a'$ just restricted to the edges of $\Pi$ is at least $2\cdot x-1$, the distance of the newly inserted points. Both matchings must also match all the $2 \cdot x - o(x)$ points of the input not incident on $\Pi$. As the minimum vertex distance between two points is $1$ this means that the cost of matching all the points not on $\Pi$ for both $\mu_a$ and $\mu_a'$ is at least $x - o(x)$. 

This implies that $c(\mu_a) + c(\mu_a') \geq 4 \cdot x - o(x)$. However, observe that there exists an optimal solution of size $x-1$ and $x$ before and after the update respectively. Hence, of the two matchings must have a cost of at least $2x - o(x)$, making it $(2- o(1))$-approximate which contradicts our initial assumption of them. Observe that picking $n$ to be arbitrarily large we can make this $o(1)$ factor arbitrarily small.

Observe that the argument trivially extends to updates consisting of point pair deletions, as we may just reverse this process.

\end{proof}

\clearpage

\section{Experimental Evaluation Plots}

\begin{figure}[H]
    \centering
    \includegraphics[width=0.4\linewidth]{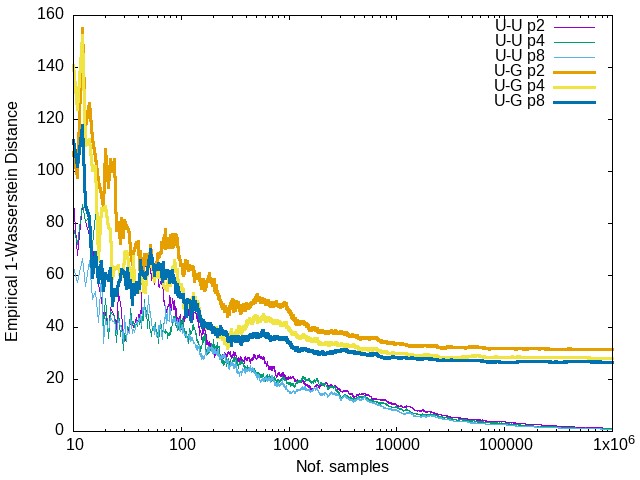}
    \includegraphics[width=0.4\linewidth]{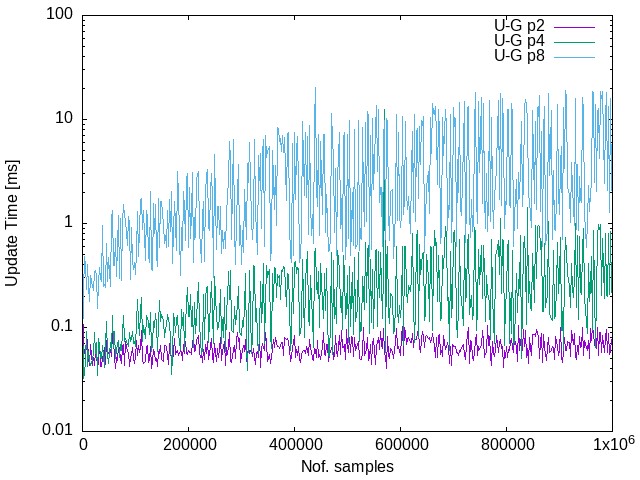}
    \caption{Convergence and update time on large datasets.}
    \label{fig:large}
\end{figure}

\begin{figure}[H]
    \centering
    \includegraphics[width=0.4\linewidth]{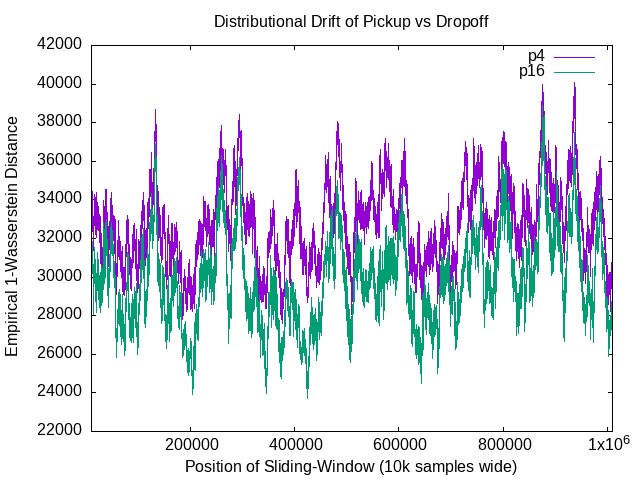}
    \includegraphics[width=0.4\linewidth]{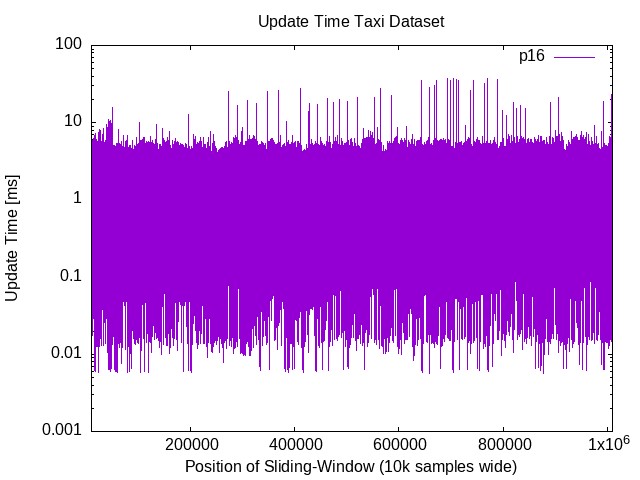}
    \caption{Drift in the Pickup-Dropoff distributions of the Taxi dataset (left) and update time (right).}
    \label{fig:drift-real}
\end{figure}

\begin{figure}[H]
    \centering
    \includegraphics[width=0.4\linewidth]{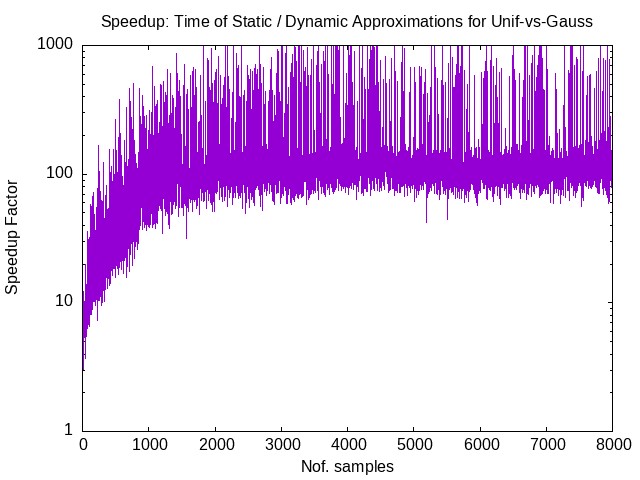}
    \includegraphics[width=0.4\linewidth]{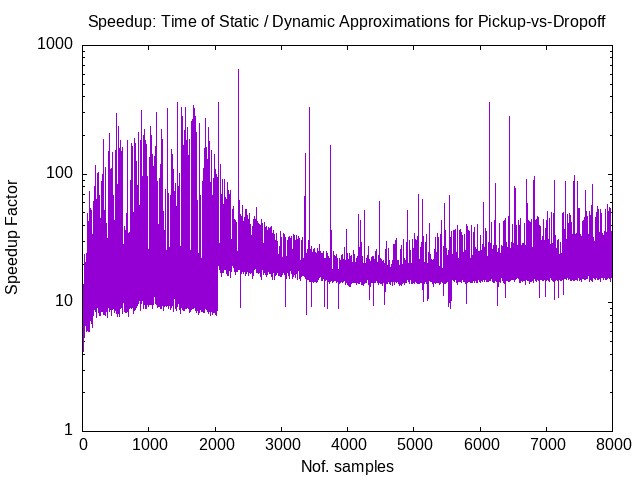}
    \caption{Speedup of dynamic algorithm over static for approximations with $p=8$.}
    \label{fig:SpeedUp}
\end{figure}

\begin{figure}[H]\centering
\includegraphics[width=0.4\linewidth]{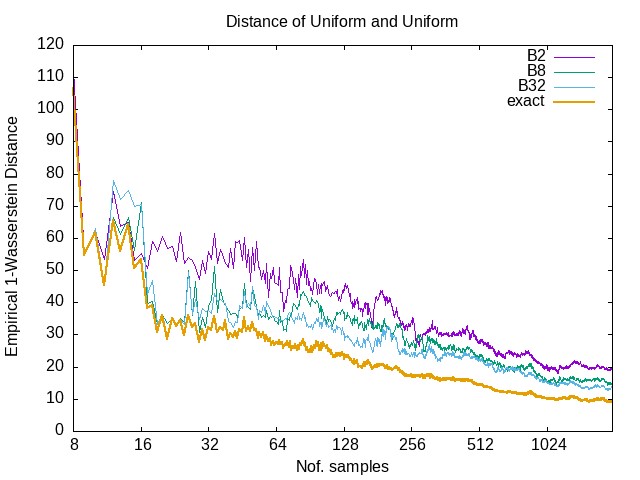}
\includegraphics[width=0.4\linewidth]{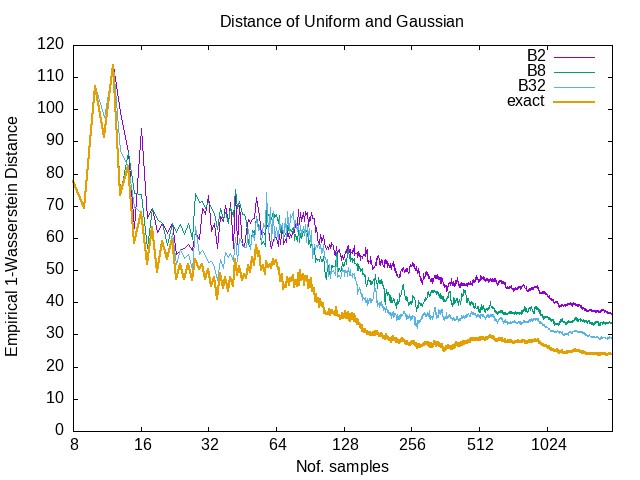}
\caption{Estimating the $1$-Wasserstein distance using exact and approximate matchings.}
\label{fig:wasser-exact-vs-apx}
\end{figure}